# Electric Field Enhancement in ε-near-zero Slabs under TM-Polarized Oblique Incidence

Salvatore Campione,[1] Domenico de Ceglia,[2,3] Maria Antonietta Vincenti,[2,3] Michael Scalora,[3] and Filippo Capolino[1,*]

[1] Department of Electrical Engineering and Computer Science, University of California Irvine, CA, 92697, USA
[2] Aegis Technologies Inc., 410 Jan Davis Dr., Huntsville, AL, 35806, USA
[3] US Army Charles M. Bowden Research Center, RDECOM, Redstone Arsenal, Huntsville, AL, 35898, USA
[*] f.capolino@uci.edu (http://capolino.eng.uci.edu)

**Abstract:** We investigate local field enhancement phenomena in subwavelength, ε-near-zero (ENZ) slabs that do not exploit Fabry-Pérot resonances. In particular, we study the linear response of engineered metamaterial slabs of finite thickness based on plasmonic nanoshells that show an ENZ band in the visible range, and naturally occurring materials (e.g., $SiO_2$) that also display ENZ properties, under oblique, TM-polarized plane wave incidence. We then introduce active gain material in engineered metamaterial slabs that adds peculiar spectral and angular features to transmission, reflection, and absorption properties, and leads to a further local field enhancement. These findings are supported by two theoretical studies: First, a simple interface between two semi-infinite media, namely free space and a generic ENZ medium; then, an ENZ slab of finite thickness, with the aim of understanding the system's behavior when varying the ENZ properties as well as the incident angle. For either case we report three distinct physical conditions for which we explain spectral and angular features that might result in strong field enhancement. The gain-assisted metamaterial implementation has the potential of triggering and enhancing low-threshold nonlinear phenomena thanks to the large local fields found at specific frequency and angular bands.

1. Introduction

In view of their potential applications, artificial composite materials that exhibit ε-near-zero (ENZ) properties [1-3] have attracted a great deal of attention. A limited list of such applications includes tunnelling of electromagnetic energy [4-5], highly directional beaming [6-12], optical nanocircuits [13], lenses [14], cloaking devices [15-16], boosting of optical nonlinearities [17] and low-threshold nonlinear effects [18-19].

A theoretical discussion of electromagnetic tunnelling through thin, ENZ channels without phase accumulation is reported in [4], where tunnelling is also implemented in anisotropic, artificial material made of wires. Theoretical predictions of artificial materials made of wire media exhibiting refractive indices less than unity have also been reported in reference [3]. An experimental demonstration of microwave tunnelling between two waveguides connected by a thin ENZ channel was discussed in [5]. The ENZ channel consisted of a planar waveguide where complementary split ring resonators were patterned on the lower surface. Experimental results were found to be in agreement with theory and numerical simulations.

The properties of ENZ materials and their applications are numerous. A notable example is their ability to radiate highly directional beams. References [8-11] have shown radiation with enhanced directivity of a transverse dipole (parallel to the interface) embedded in an ENZ slab. For example, in [8] an experimental demonstration of directional radiation through ENZ materials at microwave frequencies was carried out. The metamaterial was made of copper grids with a square lattice excited by a monopole antenna. A precise physical explanation of directive radiation of a dipole inside an ENZ material slab was established in [9-11], revealing the role of the excited leaky wave in the radiation mechanism. Based on the above discoveries, by reciprocity, one may ask if a large transverse field arises when an ENZ slab is illuminated with a plane wave close to normal incidence; however, in this paper, we will show that under plane wave illumination, in the (low-loss) ENZ condition the longitudinal field intensity (normal to the ENZ surface) is larger (up to orders of magnitude) than the transverse one. In [10], both effective medium theory (including spatial dispersion phenomena) and full-wave simulations of wire media confirmed the highly directional nature of the radiation from the low permittivity medium excited by a dipole. In [12], the use of ENZ metamaterials was proposed to tailor the phase of radiation pattern of arbitrary sources in planar layers and cylindrical shells.

ENZ materials have also been shown to prevent leakage of the optical electric displacement current in the field of metatronics [13]. In a completely different framework, stacked subwavelength hole arrays characterized either by an effective ENZ or a µ-near-zero material parameter were analyzed both theoretically and experimentally [14] in order to realize lenses. Another exotic application of ENZ metamaterials is to cloaking devices. In [15], for example, an experimental implementation of a microwave frequency cloak based on scattering cancellation technique was analyzed. The cloak was composed of an array of metallic fins embedded in a high dielectric constant environment, and was shown to cloak a dielectric cylinder by reducing 75% of the total scattering amplitude. Full-wave simulations of a cloaking device whose constituents are plasmonic nanoshells were recently implemented in [16].

Materials exhibiting ENZ properties were also recently proposed as an effective solution to stimulate nonlinear processes because of the strong field enhancement values that may be achieved when $\text{Re}(\varepsilon)$ crosses zero. The use of narrow apertures at cutoff in a plasmonic screen to design ENZ channels was used in [17] to enhance optical nonlinearities. The introduction of Kerr nonlinearities can trigger bistable and self-tunable response achieved with low threshold intensities. The effective response of a homogenized multilayered medium including Kerr nonlinearities was analyzed in [18] in the ENZ regime, showing hyperfocusing and field compensation properties, as well as propagation of nonlinear waves even if the medium linear properties would, in principle, not allow it. The field enhancement capabilities of micrometer-thick, uniaxially anisotropic ENZ slabs have been analyzed in [20], with the goal of achieving efficient second harmonic generation. The authors of [20] showed the dependence of transmissivity and field enhancement on the incident angle of a TM-polarized plane wave and on the thickness of the slab at a fixed frequency. They also investigated the role of losses in the second harmonic generation process.

These contributions notwithstanding, we believe that currently there is a need to better understand the origin of strong field enhancement effects occurring when the ENZ slab has subwavelength thickness, i.e., far from any Fabry-Pérot resonance of the ENZ etalon. In this sense, singularity-driven second and third harmonic generation in subwavelength ENZ slabs have been shown to originate from strong electric field enhancement [19]. Although transmission properties of ENZ slabs of finite thickness under oblique incidence were briefly analyzed in [12] by varying thickness and permittivity values, here we provide a thorough analysis of transmission, reflection and absorption coefficients for varying slab permittivity and incident angles, characterizing spectral and angular features for three distinct physical conditions which might result into large field enhancements inside the slab that boost nonlinear optical processes, for example.

Metamaterial implementations with ENZ properties at optical frequencies have been reported with a focus on the limitation of losses [16, 21-25]. For example, in reference [16] the authors employed plasmonic nanoshells to engineer a low-loss optical ENZ material using silver or gold. Silver-based designs yielded effective parameters with lower losses compared to gold-based designs, partly due to the fact that at optical frequencies silver is less lossy than gold [26]. In references [21-25]) the use of active ideal [21-22] and realistic gain materials embedded inside the nanoshell cores (fluorescent dye molecules in [23] and quantum dots in [24-25]) was analyzed and found to provide promising ways to design loss-compensated ENZ metamaterials.

The latter realistic scenario has motivated us to inspect field enhancement capabilities in metamaterial slabs composed of plasmonic nanoshells that exhibit an ENZ band in the visible range [23], and of materials that naturally exhibit an ENZ band (Sec. 3). Note however that alternative metamaterial implementations based on the use of low-loss plasmonic materials [27-29] may in principle be adopted and for this reason our work continues by dealing with low-loss ENZ slabs. For example, we will show that losses greatly affect the maximum achievable enhancements. For this reason we introduce an active gain material in the metamaterial design (Sec. 4). Our findings are supported by two theoretical studies (Sec. 5), namely the interface between two semi-infinite media (free space and an ENZ medium) and an ENZ slab of finite thickness in free space, which will be used in Sec. 6 to justify the behavior of the linear response observed in Sec. 4.

We demonstrate that in the absence of any active medium, the metamaterial structure may induce a field enhancement for wide frequency and angular bands. Instead, when material losses are partly compensated by introducing an active material in the nanocomposite structure we observe a much stronger field enhancement for extremely narrow frequency and angular bands. This result may for example pave the way to the development of exotic and extreme nonlinear optical phenomena.

## 2. Definition of the optical setup under consideration

We consider an optical setup composed of an ENZ slab with thickness $h$ along the $z$ direction illuminated by an obliquely incident TM-polarized plane wave, as shown in Fig. 1(a). Therefore, the incident electric field lies on the $x$-$z$ plane, i.e., $\mathbf{E}_1 = E_1(\cos\theta_i \hat{\mathbf{x}} - \sin\theta_i \hat{\mathbf{z}})e^{i\mathbf{k}_1 \cdot \mathbf{r}}$, where $E_1$ is the amplitude, $\mathbf{r} = x\hat{\mathbf{x}} + z\hat{\mathbf{z}}$ is the observation point, $\theta_i$ is the incident angle, and $\mathbf{k}_1 = k_x\hat{\mathbf{x}} + k_{z1}\hat{\mathbf{z}} = k_1(\sin\theta_i \hat{\mathbf{x}} + \cos\theta_i \hat{\mathbf{z}})$ is the wavevector, assumed here in the $x$-$z$ plane, with $k_x$ the transverse (to the $z$-axis) wavenumber, $k_{z1}$ the longitudinal wavenumber in medium 1, $k_1 = k_0\sqrt{\varepsilon_1}$, where $k_0$ is the free space wavenumber and $\varepsilon_1$ the relative permittivity in medium 1. A monochromatic, time harmonic convention $\exp(-i\omega t)$ is implicitly assumed. In the ENZ medium 2, with relative permittivity $\varepsilon_2$, we define the wavevector $\mathbf{k}_2 = k_x\hat{\mathbf{x}} + k_{z2}\hat{\mathbf{z}} = k_2(\sin\theta_t \hat{\mathbf{x}} + \cos\theta_t \hat{\mathbf{z}})$, where $k_2 = k_0\sqrt{\varepsilon_2}$ and $k_{z2} = \beta_{z2} + i\alpha_{z2}$ is the longitudinal complex wavenumber. It is useful to define the *dielectric contrast* $\hat{\varepsilon}_2 = \varepsilon_2/\varepsilon_1$ that will play an important role in the subsequent analysis. Accordingly, Snell's law is written as $\sin\theta_i = \sqrt{\hat{\varepsilon}_2}\sin\theta_t$, and $k_{z2} = \sqrt{k_2^2 - k_x^2} = k_1\sqrt{\hat{\varepsilon}_2}\cos\theta_t = k_1\sqrt{\hat{\varepsilon}_2 - \sin^2\theta_i}$, with $k_2 = k_1\sqrt{\hat{\varepsilon}_2}$.

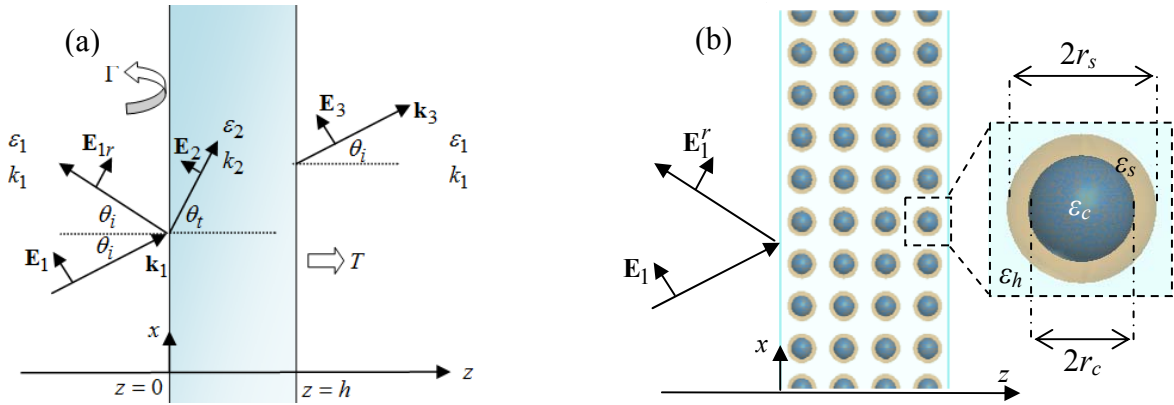

Fig. 1. (a) ENZ setup under analysis, with $|\hat{\varepsilon}_2| \ll 1$. (b) Metamaterial implementation ($x$-$z$ plane).

We define the critical incident angle $\theta_i^C$ as the angle where the vanishing longitudinal wavenumber condition $k_{z2} = k_1\sqrt{\hat{\varepsilon}_2 - \sin^2\theta_i^C} = 0$ is verified. This means that, when both the slab and the incident medium are lossless, an increase in the incident angle above $\theta_i^C$ produces a purely evanescent wave inside the slab. The critical angle may be expressed as

$$\theta_i^C = \arcsin\sqrt{\hat{\varepsilon}_2}. \qquad (1)$$

Note that the critical angle is purely real if $\hat{\varepsilon}_2 > 0$, and purely imaginary if $\hat{\varepsilon}_2 < 0$. In a more general scenario, $\theta_i^C$ is a complex number when $\hat{\varepsilon}_2$ is complex. In addition one may also define the Brewster incident angle $\theta_i^B$, the angle that yields no reflection at the interface, which will be used in what follows. The Brewster angle is characterized by the condition $k_{z2} = \hat{\varepsilon}_2 k_{z1}$, leading to

$$\theta_i^B = \arctan\sqrt{\hat{\varepsilon}_2}. \qquad (2)$$

Even though a Brewster transmission condition with zero reflection is possible only for lossless media, a minimum reflection angle may also be defined for lossy media and it depends on the imaginary part of the permittivity [30-31]. We note that for lossless media $\theta_i^B < \theta_i^C$, and for ENZ materials $\theta_i^B \approx \theta_i^C \approx \sqrt{\hat{\varepsilon}_2}$.

The most important physical parameter analyzed in this paper is the field enhancement in ENZ materials, and the conditions that lead to it. Because of the field continuity requirement at the ENZ interface, it is the $z$ component of the electric field $E_{z2}$ in the ENZ medium 2 that experiences the largest enhancement, as it will be shown next. Therefore, it is convenient to define the field intensity enhancement (FIE), evaluable at various positions $z$, as

$$\text{FIE} = |E_{z2}|^2 / |E_1|^2 \qquad (3)$$

that will be used and analyzed throughout the paper. Note that when $\theta_i = 0°$, i.e., normal incidence, both $E_{z1}$ (the z component of the total field in medium 1) and $E_{z2}$ are identically equal to zero. In principle, one may define a field enhancement also for the transverse field component $E_{x2}$ analogously to what is done in Eq. (3) for $E_{z2}$. However, we will show that in the case of (low-loss) ENZ slabs, a much larger FIE related to $E_{z2}$ is attained. Therefore, unless otherwise specified, by FIE we mean the longitudinal field intensity enhancement in Eq. (3).

## 3. Field enhancement in ENZ slabs

We consider two metamaterial slabs made of four periodic layers of plasmonic nanoshells, whose physical parameters are shown in Table I, that exhibit effective ENZ properties around 525 THz (slab 1) and 422 THz (slab 2), as previously reported in [23] via modal analysis. The nanoshell's inner core has radius $r_c$ and relative permittivity $\varepsilon_c$; the shell has outer radius $r_s$ and relative permittivity $\varepsilon_s$. The system is embedded in a homogeneous environment having relative permittivity $\varepsilon_h$, as schematically reported in Fig. 1(b). The nanoshells are periodically spaced with periodicities $a$, $b$, $c$ along the $x$, $y$ and $z$ directions, respectively. Here we consider silver (Ag) or gold (Au) shells, whose permittivity is modeled according to the Drude model $\varepsilon_s = \varepsilon_\infty - \omega_p^2 / [\omega(\omega + i\gamma)]$ ($\varepsilon_\infty = 5$, $\omega_p = 1.37 \times 10^{16}$ rad/s, $\gamma = 27.3 \times 10^{12}$ 1/s for silver; $\varepsilon_\infty = 9.5$, $\omega_p = 1.36 \times 10^{16}$ rad/s, $\gamma = 1.05 \times 10^{14}$ 1/s for gold) also reported in [23]. It has been demonstrated that for thin metallic shells, the Drude model should be modified to account for the dependence on metal thickness, surface effects and interband transitions, as for example shown in [22]. However, even a more realistic metal permittivity does not alter the qualitative analysis and the results on field enhancement in ENZ slabs shown in this paper nor does it undermine our discussion and conclusions. Metamaterial slabs are assumed to be homogeneous and to have a finite thickness $h = 4c$ along the z direction (Fig. 1). The effective permittivities of the two slabs shown in Fig. 2 (solid curves, tagged as without gain) were retrieved at normal incidence by employing the method outlined in [32] that uses reflection and transmission from a finite thickness slab computed via a full-wave simulation based on the finite element method (high frequency structure simulator, HFSS by Ansys Inc., and COMSOL Multiphysics, both in good agreement). Slab 1, made of silver shells, exhibits an effective relative permittivity $\varepsilon_{\text{eff}}$ that has a smaller imaginary part than the one of slab 2, which is made of gold shells, across the entire frequency range. The presence of a smaller amount of losses across the ENZ frequency region makes us infer that slab 1 will exhibit better performance than slab 2, as explained in the following.

Table I: Physical parameters of the nanoshells composing the metamaterial slabs.

| Slab | $\varepsilon_c$ | $\varepsilon_s$ | $\varepsilon_h$ | $r_c$ [nm] | $r_s$ [nm] | $a,b,c$ [nm] |
|---|---|---|---|---|---|---|
| 1 | 2.25 | Ag | 2.25 | 20 | 25 | 75 |
| 2 | 2.25 | Au | 2.25 | 30 | 35 | 100 |

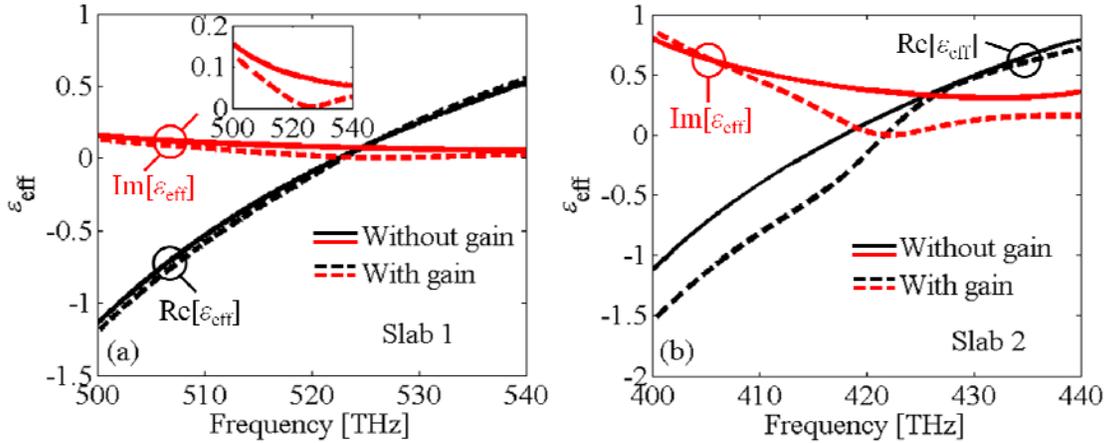

Fig. 2. Relative effective permittivity, retrieved at normal incidence, versus frequency for (a) slab 1 and (b) slab 2 in Table I, without and with gain (as explained in Sec. 4). Note the largely compensated material loss [i.e., $\text{Im}(\varepsilon_{\text{eff}}) \to 0$] near 525 THz and 422 THz, respectively, in the case with gain.

We assume slabs 1 and 2 are composed of homogeneous materials described by the functions shown in Fig. 2 (solid curves, tagged as without gain). The surrounding material in Fig.

1(a) has relative permittivity $\varepsilon_1 = 2.25$. In Fig. 3 we show the absorption coefficient $A = 1 - |T|^2 - |\Gamma|^2$, where $|T|^2$ is the transmittance, $|\Gamma|^2$ is the reflectance (equations for $T$ and $\Gamma$ are found in Sec. 5.2), and the FIE in Eq. (3) calculated at $z = 0^+$, i.e., slightly inside the slab, and at $z = h/2$. All calculations were performed by assuming TM-polarized plane wave incidence and varying frequency and incident angle for the two cases in Table I.

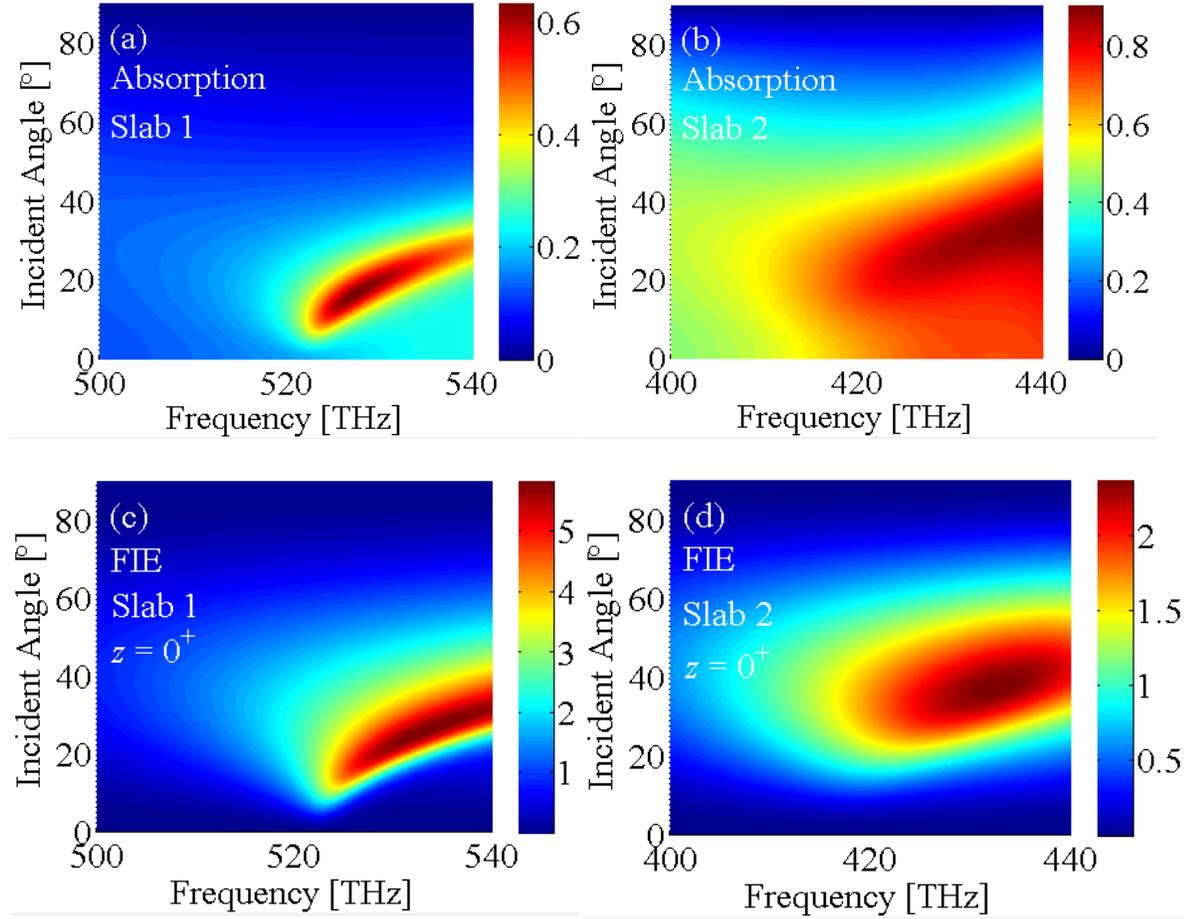

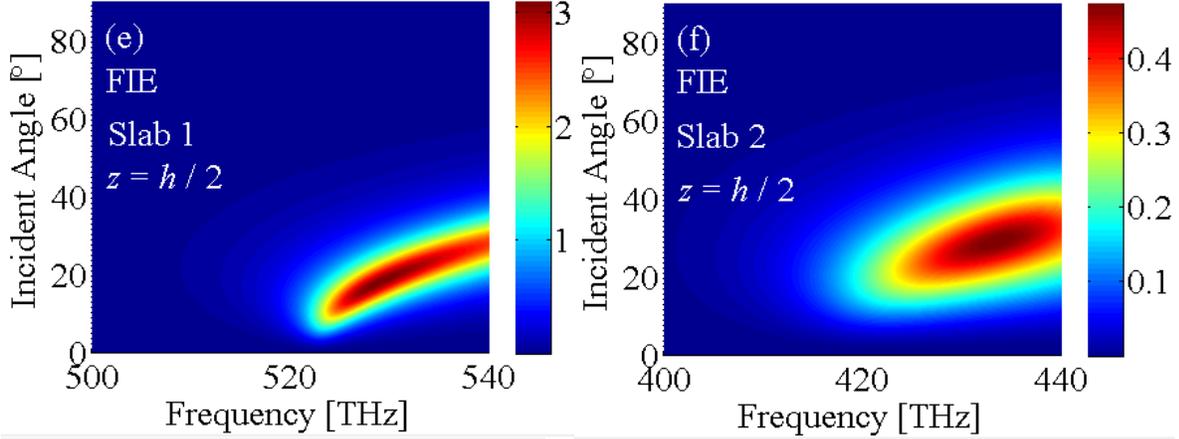

Fig. 3. Absorption coefficient $A$ as a function of frequency and incident angle for (a) slab 1 and (b) slab 2 in Table I. The FIE in Eq. (3) computed at $z = 0^+$ and $z = h/2$ as a function of frequency and incident angle for (c), (e) slab 1 and (d), (f) slab 2 in Table I. The ENZ slab thickness is $h = 4c$.

Two properties may be observed in Fig. 3: (i) absorption is larger in the case of slab 2 compared to slab 1, and has broader frequency and angular bands due to the larger $\text{Im}(\varepsilon_{\text{eff}})$; (ii) the maximum FIE in slab 1 is $\approx 2.5$ times larger than the maximum FIE achieved in slab 2 at $z = 0^+$, due to the lower $\text{Im}(\varepsilon_{\text{eff}})$ and lower losses. However, losses are still relatively large for the ENZ condition, and field intensity drops when evaluated inside the slab at $z = h/2$. These results suggest that field enhancement in an ENZ subwavelength slab is sensitive to the imaginary part of the effective permittivity. The introduction of a gain material (discussed in Sec. 4) may then boost the local field intensity and eventually support strong nonlinear phenomena. Next we excite the metamaterial slab using a TM-polarized plane wave incident at $\theta_i = 19°$ (slab 1) and $\theta_i = 30°$ (slab 2) to maximize absorption losses and field enhancement (refer to Fig. 3), and retrieve reflection, transmission and absorption coefficients (Fig. 4). From an inspection of Fig. 4 one may note a dip in the magnitude of the transmission coefficient near 522 THz for slab 1, and the absence of spectral features for slab 2, probably due to higher losses. The FIE in Eq. (3) calculated at $z = 0^+$ and $z = h/2$ is shown in Fig. 5 by varying the frequency. We observe a field enhancement for both slabs 1 and 2 when evaluated at $z = 0^+$, whereas field enhancement is still present at $z = h/2$ only for slab 1. Note that the peak tends to be located around the minimum of $\text{Im}(\varepsilon_{\text{eff}})$ (see Fig. 2). For completeness, we also show the FIE related to the $E_{x2}$ component of the field, which is lower but of a comparable order to that in Eq. (3) for $E_{z2}$. Simple calculations reveal that in the ENZ condition, the maximum attainable FIE for $E_{x2}$ is equal to 4. In this case, the presence of metal losses limits the longitudinal FIE, and as it will be shown in the next section, FIE for $E_{z2}$ can be largely enhanced by resorting to low-loss ENZ materials.

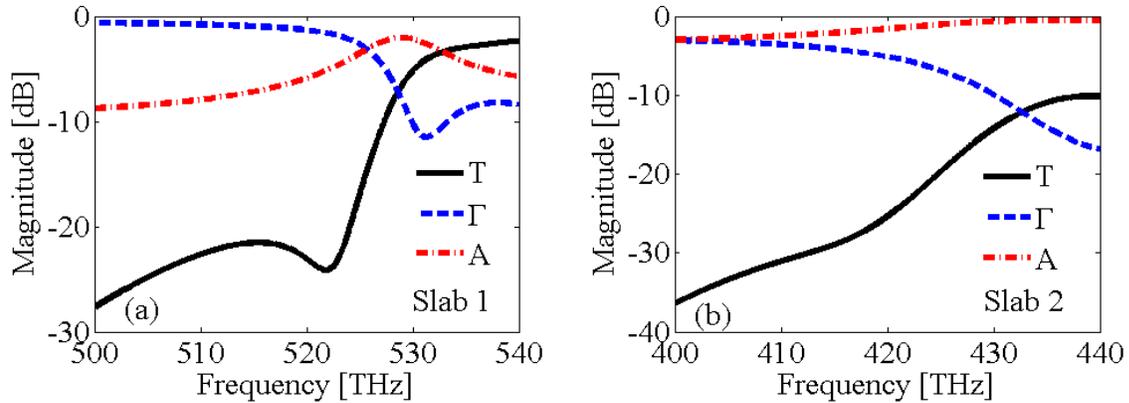

Fig. 4. Transmission, reflection and absorption coefficients as a function of frequency for (a) slab 1 and (b) slab 2 in Table I, for a TM-polarized plane wave incident at $\theta_i = 19°$ for slab 1 and $\theta_i = 30°$ for slab 2. The ENZ slab thickness is $h = 4c$.
1212

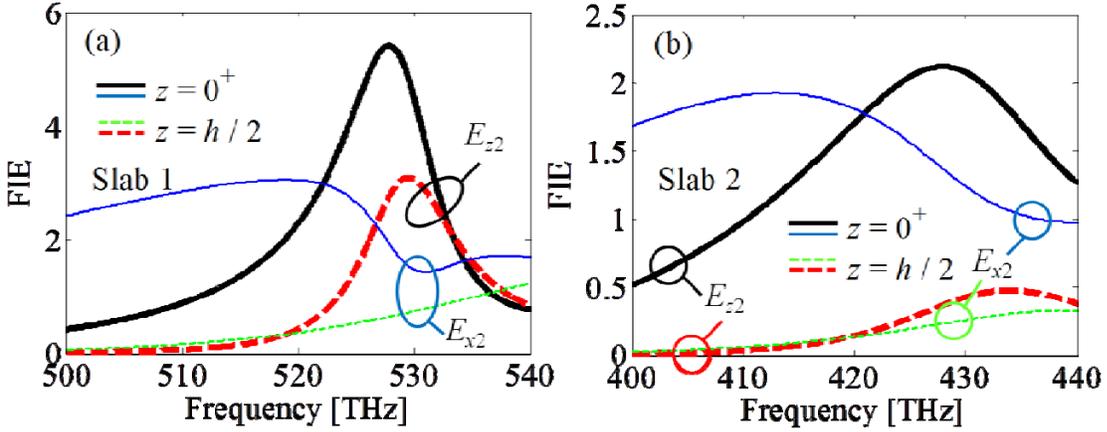

Fig. 5. The FIE in Eq. (3) computed at $z=0^+$ and $z=h/2$ for the two cases and parameters described in Fig. 4. Also, the thin blue and green lines refer to the FIE related to the $x$ component of the field $E_{x2}$. The FIE related to $E_{z2}$ is stronger, even for this case without gain.

Behavior similar to that shown in Figs. 3-5 also occurs in materials that exhibit ENZ characteristics naturally, due to bound electron resonances. For example, silicon dioxide ($SiO_2$) displays an ENZ crossing point at $\approx 37$ THz [see inset in Fig. 6(b)]. A 400 nm-thick slab surrounded by free space ($\varepsilon_1 = 1$) then exhibits large absorption and longitudinal electric field enhancement, as shown in Fig. 6. However, while absorption in the band 30-35 THz is due to increased $\mathrm{Im}(\varepsilon)$ [see inset in Fig. 6(b)], the absorption band centered at 37 THz and $60°$ is due to the condition $\mathrm{Re}(\varepsilon) \approx 0$, and yields FIE $\approx 3.5$.

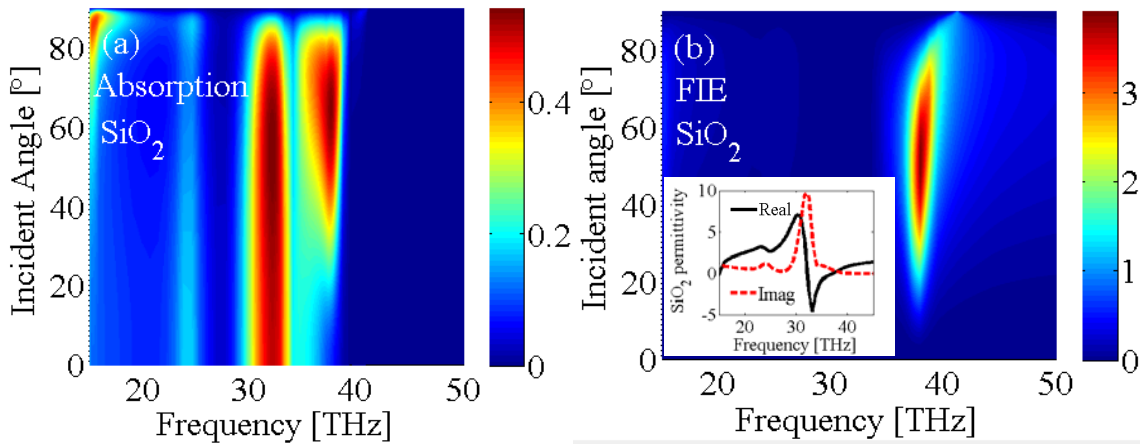

Fig. 6. (a) Absorption coefficient and (b) FIE in Eq. (3) computed at $z=0^+$ as a function of frequency and incident angle for a slab of SiO$_2$. Slab thickness is $h=400$ nm.

## 4. Super field enhancement in gain-assisted ENZ slabs

Following our discussion in the previous sections one may infer that losses and field enhancement are intimately related and that they are mutually exclusive. However, large losses are usually associated with the product $\mathrm{Im}(\varepsilon)|\mathbf{E}|^2$, and thus may arise either from large local field enhancement, or large $\mathrm{Im}(\varepsilon)$, or both. These observations motivate our analysis of a gain-assisted ENZ slab. Hence, we again consider the two composite-material slabs described in Table I, but now the nanoshells have silica-like dielectric cores that include 10 mM of Rhodamine 6G (slab 1) and Rhodamine 800 (slab 2) fluorescent dye molecules to provide gain [23]. The chosen molecular concentration is quite large and may thus impact the overall compensation due to the presence of fluorescence quenching and other non-radiative phenomena [33]. This drawback may be alleviated by using fluorescent molecules that exhibit larger emission cross section compared to those considered above, or by using alternative approaches, such as the inclusion of quantum dots [24, 34]. The effective permittivity of a sample composed of 4 layers of such nanoshells (slab thickness is $h=4c$) is reported in Fig. 2 for normal incidence (dashed curves, tagged as with gain). We note that $\mathrm{Re}(\varepsilon_\mathrm{eff})$ is affected only slightly by the introduction of gain, as may be ascertained by comparing solid and dashed curves in Fig. 2. A careful comparison of such curves reveals that $\mathrm{Im}(\varepsilon_\mathrm{eff})$ is reduced considerably near the zero-crossing point (525 THz for slab 1 and 422 THz for slab 2).

In Fig. 7 we show the gain-assisted absorption coefficient $A$ and the FIE in Eq. (3) computed at $z=0^+$ and $z=h/2$ under TM-polarized plane wave incidence as a function of frequency and incident angle for the two cases in Table I. The surrounding material in Fig. 1(a) has relative permittivity $\varepsilon_1 = 2.25$. A comparison of Figs. 3 and 7 reveals that the angle of maximum FIE decreases as $\text{Im}(\varepsilon_{\text{eff}})$ decreases, fact that may be ascertained by observing that the FIE reaches $\approx 35$ for slab 1 ($\varepsilon_{\text{eff}} \approx 0.12 + i3\times10^{-3}$ at about 526 THz) and $\approx 180$ for slab 2 ($\varepsilon_{\text{eff}} \approx 0.03 + i10^{-4}$ at about 422 THz). When compared to the maps in Fig. 3, the FIE remains nearly constant inside both gain-assisted slabs. In general, adding gain to the metamaterial response lowers the damping of the system and thus the losses. Nevertheless, the simultaneous availability of a $\text{Re}(\varepsilon_{\text{eff}}) = 0$ crossing point and smaller imaginary part leads to a large enhancement of the longitudinal field at low incident angles, a condition in which the system experiences absorption rates similar to absorption rates found in the absence of gain. The high absorption band centered at about 435 THz in Fig. 7(b) is due to the first-order Fabry-Pérot resonance in slab 2 (where the real part of the effective permittivity is about 0.75). Fabry-Pérot resonators require a π-phase accumulation through the slab, so that larger thicknesses $h$ or larger permittivity values should be considered for the etalon. This resonant feature occurs in a similar way also for TE-polarized incident illumination.

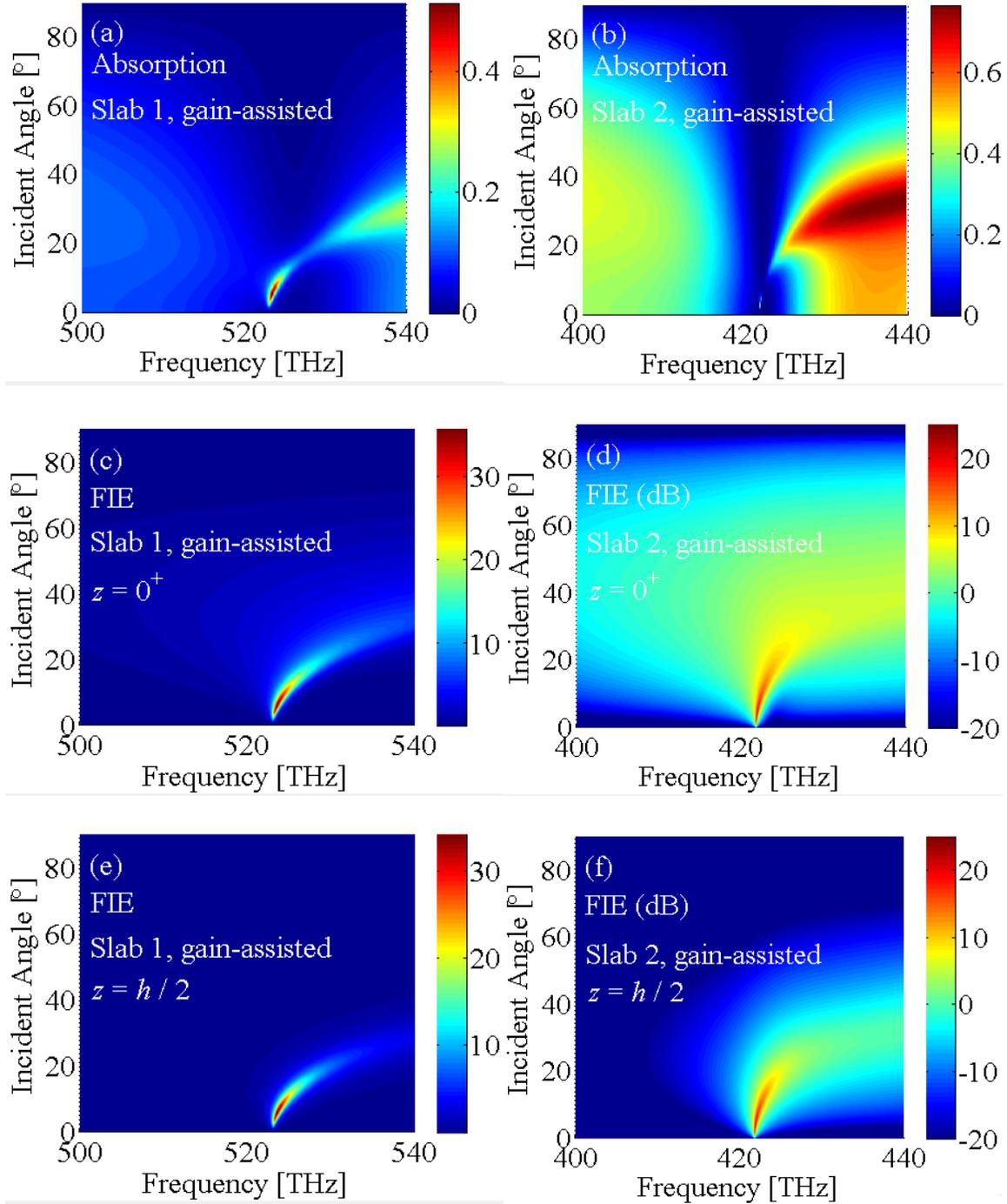

Fig. 7. As in Fig. 3, for the two gain-assisted metamaterials slabs in Table I. Note the logarithmic scale $[10\log_{10}(\text{FIE})]$ is used for slab 2, in (d), (f), due to the large FIE values.

We then excite the two gain-assisted slabs with TM-polarized plane waves incident at $\theta_i = 6°$ for slab 1 and $\theta_i = 3°$ for slab 2, i.e., at the angles that maximize absorption losses and field enhancement (refer to Fig. 7), and retrieve reflection, transmission and absorption coefficients (Fig. 8). In both cases, around the $\varepsilon \approx 0$ frequency range, the transmission coefficient is strongly *asymmetric* (i.e., a dip followed by a peak for increasing frequency) as absorption increases, conditions that are not observed in the absence of gain (Fig. 4). We will discuss these features in Sec. 6.

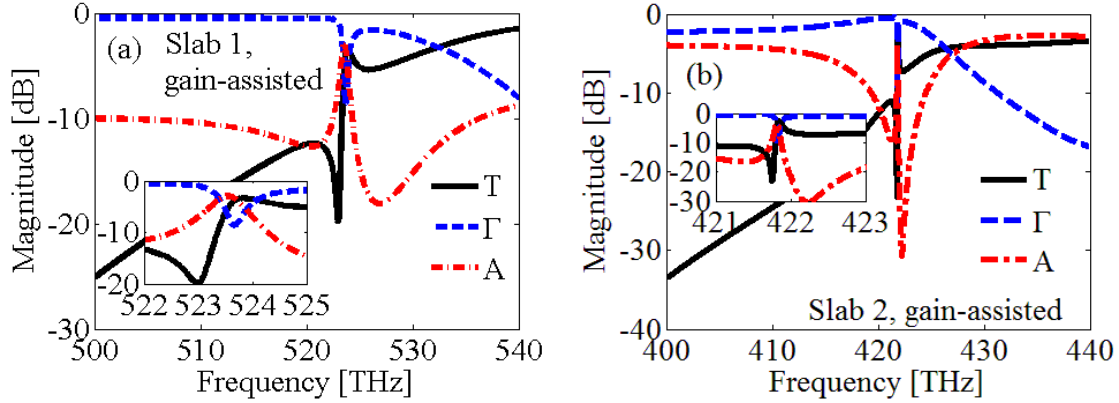

Fig. 8. Magnitude of transmission, reflection and absorption coefficients as functions of frequency for gain-assisted (a) slab 1 and (b) slab 2 in Table I, assuming a TM-polarized plane wave incident at $\theta_i = 6°$ for slab 1 and $\theta_i = 3°$ for slab 2. The ENZ slab thickness is $h = 4c$.

The FIE in Eq. (3) computed at $z = 0^+$ and $z = h/2$ is shown in Fig. 9, where a FIE $\approx 35$ in the case of slab 1, and $\approx 180$ in the case of slab 2 is predicted. Compared to Fig. 5, we note that this enhancement remains nearly constant when evaluated inside the slab, especially for slab 2 because it exhibits a smaller $\text{Im}(\varepsilon_{\text{eff}})$ with respect to slab 1. The peak does not move as in Fig. 5 because here the conditions $\text{Re}(\varepsilon_{\text{eff}}) \approx 0$ and $\text{Im}(\varepsilon_{\text{eff}}) \approx 0$ occur at nearly the same frequency (Fig. 2). The enhancements observed here are promising for applications to exotic and extreme nonlinear optical phenomena. We also show that the FIE related to the *x* component of the field, $E_{x2}$, is now negligible (especially for Slab 2) with respect to the one achieved via the super-enhancement of $E_{z2}$ due to the low-loss ENZ condition. Indeed, Fig. 9 shows that in this case the FIE for $E_{z2}$ is much higher than what shown in Fig. 5. This explains why the main focus of this paper is the longitudinal FIE in Eq. (3).

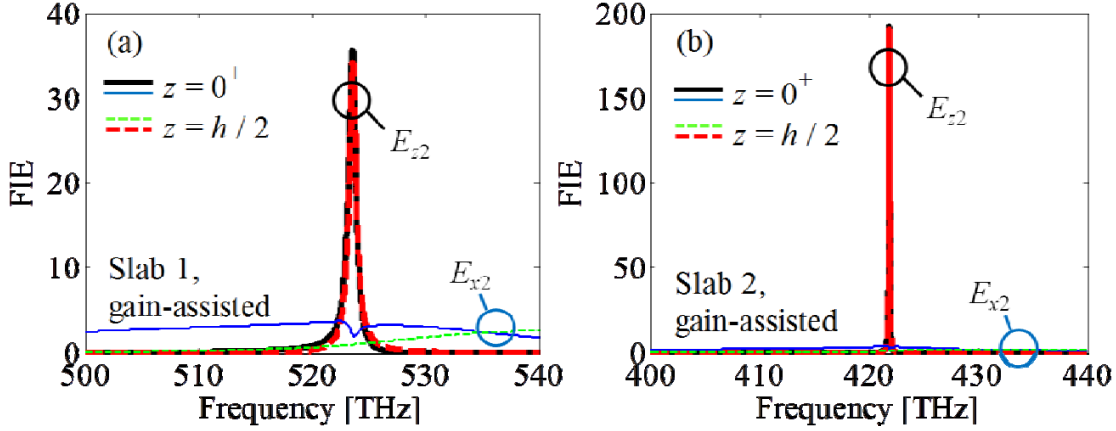

Fig. 9. The FIE in Eq. (3) computed at $z = 0^+$ and $z = h/2$ for the two cases and parameters described in Fig. 8. Also, the blue and green thin lines refer to the FIE related to the $x$ component of the field $E_{x2}$. The FIE related to $E_{z2}$ is much stronger than the one observed in Fig. 5, due to the low-loss ENZ condition.

## 5. Analytical model

The field enhancement observed in ENZ metamaterial slabs (Figs. 3-5), which is further improved in gain-assisted systems (Figs. 7-9), may be adopted for efficient second and third harmonic generation schemes or other applications where the presence of large fields is critical[35]. The next subsections will be devoted to the introduction of the model required to explain the field enhancement associated to gain-assisted ENZ slabs observed in Sec. 4. This model is an effective tool that may be used to predict field enhancement, spectra features and the effect of losses.

### 5.1. Interface between two half spaces

We begin by analyzing the case of an interface between two half spaces, obtained by assuming $h \to +\infty$ (see Fig. 1). Following the formulation in reference [36], the incident and transmitted fields into the ENZ medium are $\mathbf{E}_1 = E_1 (\cos\theta_i \hat{\mathbf{x}} - \sin\theta_i \hat{\mathbf{z}}) e^{i\mathbf{k}_1 \cdot \mathbf{r}}$, and $\mathbf{E}_2 = E_1 (\cos\theta_i / \cos\theta_t) T_0 (\cos\theta_t \hat{\mathbf{x}} - \sin\theta_t \hat{\mathbf{z}}) e^{i\mathbf{k}_2 \cdot \mathbf{r}}$, where the transmission coefficient $T_0$ is defined as $T_0 = E_{x2} / E_{x1}$, using the transverse field components. Assuming $E_{x1}^r$ is the $x$ component of the reflected field, the reflection coefficient is defined as $\Gamma_0 = E_{x1}^r / E_{x1}$. Both $\Gamma_0(\hat{\varepsilon}_2, \theta_i)$ and $T_0(\hat{\varepsilon}_2, \theta_i)$ are functions of $\hat{\varepsilon}_2$ and $\theta_i$, and are given by

$$\Gamma_0(\hat{\varepsilon}_2,\theta_i) = \frac{k_{z2} - \hat{\varepsilon}_2 k_{z1}}{k_{z2} + \hat{\varepsilon}_2 k_{z1}} \quad T_0(\hat{\varepsilon}_2,\theta_i) = \frac{2k_{z2}}{k_{z2} + \hat{\varepsilon}_2 k_{z1}}. \quad (4)$$

The values of $\Gamma_0(\hat{\varepsilon}_2,\theta_i)$, $T_0(\hat{\varepsilon}_2,\theta_i)$, and of the $z$ component of the field $E_{z2} = -E_{x2} \tan\theta_t$ are examined in what follows under certain parameter and incidence conditions. For what concerns the reflection coefficient in the general lossy case, only two sets of parameters lead to total reflection (i.e., $|\Gamma_0(\hat{\varepsilon}_2,\theta_i)| = 1$): (i) $\hat{\varepsilon}_2 = 0$ and $\theta_i \neq 0$, treated in Sec. 5.1a and referred to as epsilon equal zero (EEZ) condition; (ii) $\hat{\varepsilon}_2 \neq 0$ and $\theta_i = \theta_i^C$, discussed in Sec. 5.1c and referred to as critical angle condition (CAC). If we limit the analysis to the lossless case, then the sets of parameters that produce total reflection may be summarized as follows:

$$\begin{aligned}
\hat{\varepsilon}_2 < 0, \; \theta_i \neq 0 &\rightarrow |\Gamma| = 1 \\
\hat{\varepsilon}_2 = 0, \; \theta_i \neq 0 &\rightarrow \Gamma = 1 \\
\hat{\varepsilon}_2 > 0, \; \theta_i = \theta_i^C &\rightarrow \Gamma = -1 \\
\hat{\varepsilon}_2 > 0, \; \theta_i > \theta_i^C &\rightarrow |\Gamma| = 1
\end{aligned} \quad (5)$$

The $z$ component of the field evaluated at the boundary ($z = 0^+$) may be also computed as $E_{z2} = E_{z1} / \hat{\varepsilon}_2$, where $E_{z1}$ is the $z$ component of the total field in medium 1 ($z = 0^-$), accounting also for reflection. Three important limiting conditions are investigated in subsections 5.1a-c stressing the enhancement of the $E_{z2}$ field component.

### a. Epsilon equal zero condition

The epsilon equal zero (EEZ) condition is defined by $\hat{\varepsilon}_2 = 0$ for any $\theta_i \neq 0$, and it results in $\Gamma_0(0,\theta_i) = 1$ and $T_0(0,\theta_i) = 2$. This condition thus leads to total reflection. Note that since $k_{z2} = k_1\sqrt{\hat{\varepsilon}_2 - \sin^2\theta_i}$, when $\hat{\varepsilon}_2 \to 0$ one then has $k_{z2} \to ik_1\sin\theta_i$. The square root sign satisfies the boundary conditions at $z \to \infty$ [37], i.e., decaying amplitude for increasing $z$. Indeed, given the expression for the field $E_{z2}$ in Sec. 5.1, one observes that when performing the limit for $\hat{\varepsilon}_2 \to 0$, $E_{z2}$ is given by

$$E_{z2}(x,z) = i2\cos\theta_i E_1 e^{k_1\sin\theta_i(ix-z)}. \quad (6)$$

Expression (6) suggests that light propagation in the ENZ medium is forbidden. Moreover, $E_{z2}(x,z) \to i2E_1$ when $\theta_i \to 0$, in contrast to what happens for a finite thickness slab in Sec. 5.2. One may also look at the continuity of the component of displacement field normal to the $z=0$ boundary $E_{z1} = \hat{\varepsilon}_2 E_{z2}$. This condition seems to suggest that $E_{z2} \to \infty$ as $\hat{\varepsilon}_2 \to 0$. However, we note that at $z=0^-$ one has that $E_{z1} = -E_1\sin\theta_i\left[1-\Gamma_0(0,\theta_i)\right] \to 0$ because $\left[1-\Gamma_0(0,\theta_i)\right] \to 0$ as $\hat{\varepsilon}_2$, and the value of $E_{z2}$ at $z=0^+$ is obtained by a limiting operation, leading to (6).

### b. Total transmission condition

Total transmission condition (TTC) is defined by $|\Gamma_0(\hat{\varepsilon}_2, \theta_i)| = 0$. The condition that leads to TTC is $\hat{\varepsilon}_2 \neq 0$ and $\theta_i = \theta_i^B$ so that $\Gamma_0(\hat{\varepsilon}_2, \theta_i^B) = 0$ and $T_0(\hat{\varepsilon}_2, \theta_i^B) = 1$. The $z$ component of the field $E_{z2}$ is

$$E_{z2}(x,z) = -E_1 \frac{1}{\sqrt{\hat{\varepsilon}_2(1+\hat{\varepsilon}_2)}} e^{ik_1 \frac{\sqrt{\hat{\varepsilon}_2}}{\sqrt{1+\hat{\varepsilon}_2}}(x+z\sqrt{\hat{\varepsilon}_2})}. \qquad (7)$$

which in general is a wave traveling while decaying along $x$ and $z$. The field in (7) has an inverse dependence on $\hat{\varepsilon}_2$, a relationship that is important for the attainment of large field enhancement.

If we now assume that $\hat{\varepsilon}_2 \to 0$, Eq. (2) imposes that $\theta_i = \theta_i^B \to 0$ and Eq. (7) predicts infinite field values, in contrast with Eq. (6). This apparent ambiguity on the $E_{z2}$ value for $(\hat{\varepsilon}_2, \theta_i) \to (0,0)$ critically depends on the path selected on the $(\hat{\varepsilon}_2, \theta_i)$ space to approach $(0,0)$. If $\hat{\varepsilon}_2 \to 0$ at the TTC, $\theta_i$ goes to 0 as $\arctan(\sqrt{\hat{\varepsilon}_2})$. At $z = 0^-$ one has that $E_{z1} = -E_1 \sin\theta_i \left[1 - \Gamma_0(\hat{\varepsilon}_2, \theta_i^B)\right] = -E_1 \sin\theta_i$ goes to 0 as $\sqrt{\hat{\varepsilon}_2}$ so that, from the field continuity equation, $E_{z2} = E_{z1}/\hat{\varepsilon}_2$ at $z = 0^+$ goes to infinity as $1/\sqrt{\hat{\varepsilon}_2}$, as shown in Eq. (7).

### c. Critical angle condition

As mentioned at the beginning of Sec. 5.1, the critical angle condition (CAC) occurs when $\hat{\varepsilon}_2 \neq 0$ and $\theta_i = \theta_i^C$, leading to $\Gamma_0(\hat{\varepsilon}_2, \theta_i^C) = -1$ and $T_0(\hat{\varepsilon}_2, \theta_i^C) = 0$. The transmission coefficient goes to zero as $\cos\theta_t$ in correspondence of the critical angle, thus $\lim_{\theta_i \to \theta_i^C} T_0(\hat{\varepsilon}_2, \theta_i) = 2\cos\theta_t / (\cos\theta_i \sqrt{\hat{\varepsilon}_2})$. This result is used to compute the $z$ component of the electric field in medium 2, given by

$$E_{z2}(x,z) = -\frac{2}{\sqrt{\hat{\varepsilon}_2}} E_1 e^{ik_1\sqrt{\hat{\varepsilon}_2}x}, \qquad (8)$$

which in general represents a wave traveling while decaying along $x$. Similarly to TTC, the field in (8) depends inversely on $\hat{\varepsilon}_2$, thus emphasizing the importance of ENZ capabilities of medium 2 for field enhancements. If we now consider that $\hat{\varepsilon}_2 \to 0$, Eq. (8) predicts an infinite value for the field, in contrast to Eq. (6). Similarly to what happens at the TTC, if $\hat{\varepsilon}_2 \to 0$ at the CAC, $\theta_i$ goes to 0 as $\arcsin(\sqrt{\hat{\varepsilon}_2})$. At $z = 0^-$ one has that $E_{z1} = -E_1\sin\theta_i\left[1-\Gamma_0(\hat{\varepsilon}_2,\theta_i^C)\right] = -2E_1\sin\theta_i$ goes to 0 as $\sqrt{\hat{\varepsilon}_2}$, so that $E_{z2} = E_{z1}/\hat{\varepsilon}_2$ at $z = 0^+$ goes to infinity as $1/\sqrt{\hat{\varepsilon}_2}$, as shown in Eq. (8).

    *d. Representative example 1: Analysis by varying the permittivity of the ENZ slab*

We now suppose to have a TM-polarized plane wave incident at $\theta_i = 20°$ at 400 THz. Transmission, reflection, and the FIE in Eq. (3) calculated at $z = 0^+$ as a function of the dielectric contrast ranging in the interval $-0.1 < \hat{\varepsilon}_2 < 0.3$ are shown in Fig. 10. When $\hat{\varepsilon}_2 = 0$, $\theta_i \neq 0$, $\Gamma_0(0,\theta_i) = 1$ and $T_0(0,\theta_i) = 2 \equiv 6\text{ dB}$, as predicted in Sec. 5.1a, and $\text{FIE} \approx 3.5$, as expected from Eq. (6). The CAC at $\theta_i^C = 20°$ is verified when $\hat{\varepsilon}_2 \approx 0.117$ [Eq. (1)], and $|\Gamma_0(\hat{\varepsilon}_2,\theta_i^C)| = 1$ and $T_0(\hat{\varepsilon}_2,\theta_i^C) = 0$, as predicted in Sec. 5.1c; interestingly, the FIE in Eq. (3) peaks at $\theta_i^C$ ($\text{FIE} \approx 34$) as dictated by Eq. (8). When $\hat{\varepsilon}_2 \approx 0.1325$ the TTC at the Brewster angle corresponds to $\theta_i^B = 20°$ [Eq. (2)], and one has $\Gamma_0\left(\hat{\varepsilon}_2,\theta_i^B\right)=0$ and $T_0\left(\hat{\varepsilon}_2,\theta_i^B\right)=1$, as predicted in Sec. 5.1b; the FIE in Eq. (3) is about 6.6 as in Eq. (7). Note also that in view of Eq. (5), when $\hat{\varepsilon}_2 < 0$ or $0 < \hat{\varepsilon}_2 < 0.117$ (i.e., $\theta_i > \theta_i^C$) the reflection coefficient has unit magnitude because we are considering a lossless framework.

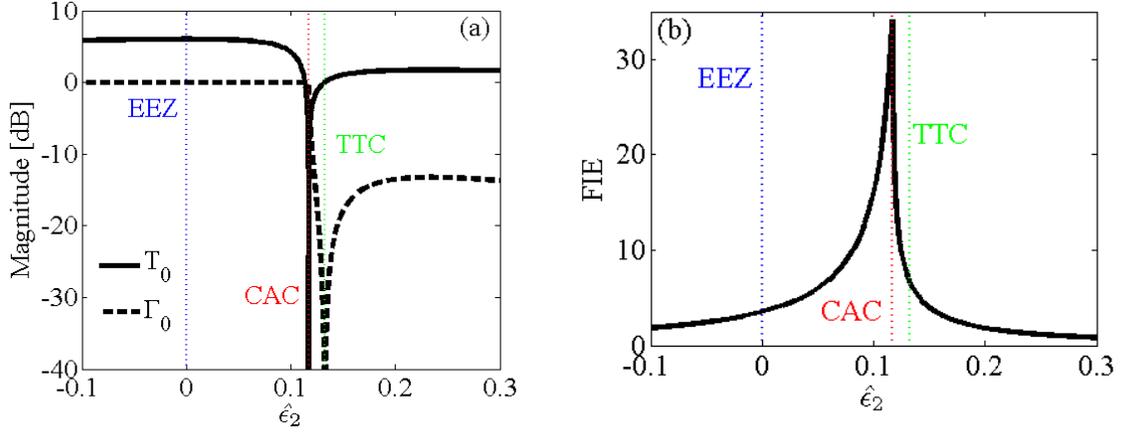

Fig. 10. (a) Transmission and reflection, and (b) the FIE in Eq. (3) computed at $z=0^+$ as a function of the dielectric contrast $\hat{\varepsilon}_2$ assuming a TM-polarized plane wave incident at $\theta_i = 20°$ at 400 THz, in the case of the interface between two half spaces.

*e. Representative example 2: Analysis by varying the angle of incidence*

To verify all the conditions in Eq. (5) in the lossless case, we consider now a TM-polarized plane wave incident with an angle in the range $0° < \theta_i < 30°$ at 400 THz for the same setup described in Fig. 10, for different values of the dielectric constrast $\hat{\varepsilon}_2$. In Fig. 11 we show reflection and the FIE in Eq. (3) calculated at $z=0^+$ as a function of the incident angle $\theta_i$. When $\hat{\varepsilon}_2 = -0.1$, one observes $\left|\Gamma_0\left(\hat{\varepsilon}_2,\theta_i\right)\right|=1$ for any $\theta_i$, and the FIE is quite limited. When $\hat{\varepsilon}_2 = 0$, one observes again $\left|\Gamma_0\left(\hat{\varepsilon}_2,\theta_i\right)\right|=1$ for any $\theta_i$, and the FIE follows a $\cos\theta_i$ envelope for $\theta_i \neq 0°$ as dictated by Eq. (6). Finally, when $\hat{\varepsilon}_2 = 0.117$ (same case as in Fig. 10 for which $\theta_i^C = 20°$) and when $\hat{\varepsilon}_2 = 0.05$ (for which $\theta_i^C = 12.92°$), one observes that $|\Gamma_0(\hat{\varepsilon}_2, \theta_i)| = 1$ for $\theta_i \geq \theta_i^C$; we note that maximum FIE occurs in correspondence of the critical angle, as observed in Fig. 10. Notably, the smaller the parameter $\hat{\varepsilon}_2$ is, the larger the FIE will be (Fig. 11), as previously mentioned in Secs. 5.1b and 5.1c, where we demonstrated the longitudinal field component to have an inverse dependence on $\hat{\varepsilon}_2$ [Eqs. (7) and (8)]. The dip in the reflection coefficient in Fig. 11 for the cases with $\hat{\varepsilon}_2 = 0.117$ and $\hat{\varepsilon}_2 = 0.05$ is in correspondence of the Brewster angles $\theta_i^B = 18.88°$ and $\theta_i^B = 12.6°$, respectively.

One may conclude that the CAC gives the maximum longitudinal field intensity enhancement for interfaces with ENZ media. Indeed, referring to (6)-(8) for $z = 0^+$, one can observe that $|E_{z2}^{\text{EEZ}}| = 2\cos\theta_i E_1$, $|E_{z2}^{\text{TTC}}| = (1/\sqrt{\hat{\varepsilon}_2})\sin\theta_t E_1 = (1/\hat{\varepsilon}_2)\sin\theta_i^B E_1$, and $|E_{z2}^{\text{CAC}}| = (2/\sqrt{\hat{\varepsilon}_2})E_1$, that imply $|E_{z2}^{\text{TTC}}|, |E_{z2}^{\text{CAC}}| > |E_{z2}^{\text{EEZ}}|$ in ENZ condition. If we now analyze the ratio $|E_{z2}^{\text{CAC}}|/|E_{z2}^{\text{TTC}}|$ in ENZ condition then $|E_{z2}^{\text{CAC}}|/|E_{z2}^{\text{TTC}}| = (2\sqrt{\hat{\varepsilon}_2})/\sin\theta_i^B = 2\sqrt{1+\hat{\varepsilon}_2} \approx 2$, which shows that the $z$ component of the field at the CAC condition is the largest. In summary, the physical properties of the interface between two half spaces, one of which exhibits near-zero permittivity, are highly dependent on the illumination incident angle and the value of the ENZ permittivity itself.

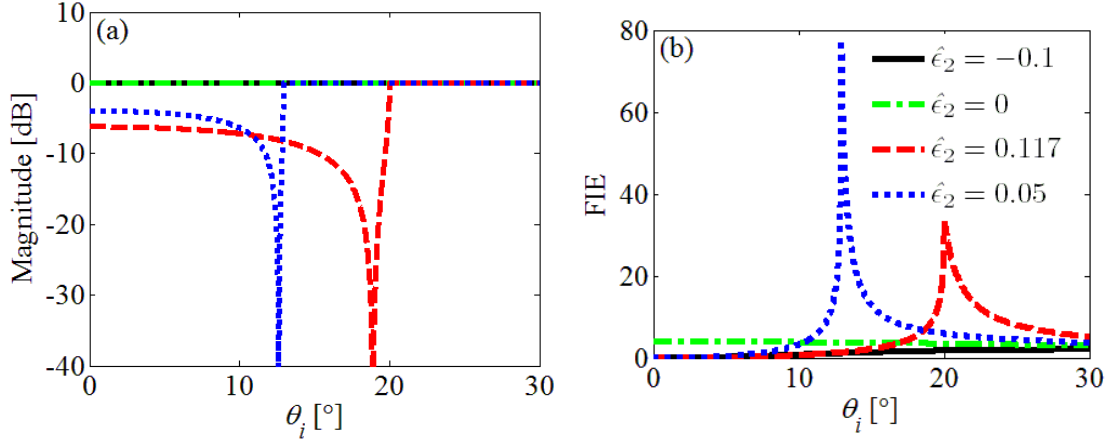

Fig. 11. (a) Magnitude of reflection coefficient and (b) the FIE in Eq. (3) computed at $z = 0^+$ as a function of the incident angle $\theta_i$ for different values of the dielectric contrast $\hat{\varepsilon}_2$, assuming a TM-polarized plane wave incidence at 400 THz, in the case of the interface between two half spaces.

### 5.2. Finite thickness slab with ENZ properties

We now analyze a slab with finite thickness $h$ (see Fig. 1). In this case, the transmitted field is $\mathbf{E}_3 = E_1 T \left( \cos\theta_i \hat{\mathbf{x}} - \sin\theta_i \hat{\mathbf{z}} \right) e^{i\mathbf{k}_3 \cdot \mathbf{r}}$, and both reflection and transmission coefficients are functions of $\hat{\varepsilon}_2$ and $\theta_i$. For simplicity it is useful to define $s_h = \sin(k_{z2} h)$ and $c_h = \cos(k_{z2} h)$. The reflection and transmission coefficients are then given by [36-37]

$$\Gamma(\hat{\varepsilon}_2, \theta_i) = \frac{-i\left(k_{z2}^2 - k_{z1}^2 \hat{\varepsilon}_2^2\right) s_h}{2 k_{z1} k_{z2} \hat{\varepsilon}_2 c_h - i\left(k_{z2}^2 + k_{z1}^2 \hat{\varepsilon}_2^2\right) s_h},$$

$$T(\hat{\varepsilon}_2, \theta_i) = \frac{2 k_{z1} k_{z2} \hat{\varepsilon}_2}{2 k_{z1} k_{z2} \hat{\varepsilon}_2 c_h - i\left(k_{z2}^2 + k_{z1}^2 \hat{\varepsilon}_2^2\right) s_h}.$$

(9)

The field inside the slab may be expressed as a superposition of forward and backward waves as

$$\mathbf{E}_2(x,z) = E_1 e^{i k_2 x \sin\theta_t} \frac{\cos\theta_i}{\cos\theta_t} \times$$
$$\times \left[ A_f \left( \cos\theta_t \hat{\mathbf{x}} - \sin\theta_t \hat{\mathbf{z}} \right) e^{i k_2 z \cos\theta_t} + A_b \left( \cos\theta_t \hat{\mathbf{x}} + \sin\theta_t \hat{\mathbf{z}} \right) e^{-i k_2 z \cos\theta_t} \right]$$

(10)

where

$$A_f = \frac{1-\Gamma_0\Gamma}{1-\Gamma_0}, \quad A_b = \frac{\Gamma-\Gamma_0}{1-\Gamma_0}, \quad (11)$$

with $\Gamma_0$ as in Eq. (4) and $\Gamma$ as in Eq. (9).

Expressions (9)-(11) will be used in some of the next subsections to estimate the value of transmission, reflection and field enhancement in correspondence of EEZ, TTC and CAC conditions.

*a. Epsilon equal zero condition*

The set of parameters that leads to EEZ is $\hat{\varepsilon}_2 = 0$ and $\theta_i \neq 0$ for which $\Gamma(0,\theta_i) = \lim_{\hat{\varepsilon}_2 \to 0} \Gamma(\hat{\varepsilon}_2,\theta_i) = 1$ and $T(0,\theta_i) = \lim_{\hat{\varepsilon}_2 \to 0} T(\hat{\varepsilon}_2,\theta_i) = 0$. Looking at the expression of the $z$ component of the field $E_{z2}$ given in (10) one may observe that when performing the limit for $\hat{\varepsilon}_2 \to 0$, $E_{z2}$ is

$$E_{z2}(x,z) = -2\cos\theta_i E_1 e^{ik_1 x \sin\theta_i} \frac{\cos[ik_1 \sin\theta_i (z-h)]}{\sin(ik_1 h \sin\theta_i)}. \quad (12)$$

Alternatively and similarly to what already described in Sec. 5.1a, the condition at the $z=0$ boundary $E_{z1} = \hat{\varepsilon}_2 E_{z2}$ seems to imply that the longitudinal component of the electric field $E_{z2}$ inside the ENZ slab becomes singular when $\hat{\varepsilon}_2 \to 0$. However, we note that at $z = 0^-$ one has that $E_{z1} = -E_1 \sin\theta_i [1-\Gamma_0(0,\theta_i)] \to 0$ because $[1-\Gamma_0(0,\theta_i)] \to 0$ as $\hat{\varepsilon}_2$, and the value of $E_{z2}$ at $z = 0^+$ is obtained by a limiting operation, leading to (12). The field in Eq. (12) is a function of the thickness $h$ and the incident angle $\theta_i$, and tends to infinity as either $h$ or $\theta_i$ (or both) tends to zero, as it will be shown next. This behavior is dramatically different from the situation described for an interface between two half spaces in Sec. 5.1a, where the field was limited as in Eq. (6). Under EEZ conditions, evanescent waves are excited even for very small incident angles above the critical angle. While for a single interface evanescent waves only attenuate in the EEZ medium, the presence of a second interface at $z = h$ allows multiple reflections of these waves that add in phase inside the EEZ slab and lead to the singular behavior in Eq. (12). Indeed the field in Eq. (12) is a superposition of two evanescent waves, represented by the term $\cos[ik_1 \sin\theta_i (z-h)]$, whose amplitude $\propto 1/\sin(ik_1 h \sin\theta_i)$ is inversely proportional to both $\theta_i$ and $h$. Moreover, we point out that Eq. (12) tends to Eq. (6) when $h \to +\infty$, whereas at $z = 0^+$ it can be rewritten as $E_{z2} = E_{z1}/\hat{\varepsilon}_2 = 2i\cos\theta_i E_1 e^{ik_1 x \sin\theta_i} \coth(k_1 h \sin\theta_i)$. In the approximation of small $\theta_i$, it follows that $E_{z2} \sim 2iE_1 e^{ik_1 x \theta_i}/(k_1 h \theta_i)$, whereas for small $h$ it is $E_{z2} \sim 2i\cos\theta_i E_1 e^{ik_1 x \sin\theta_i}/(k_1 h \sin\theta_i)$. In either case, we observe a singular FIE with respect to both small thickness and small incident angle. This is a general condition valid for ENZ slabs, and it means that if we empirically assume that the slabs shown in Sec. 3 and Sec. 4 exhibit $\hat{\varepsilon}_2 = 0$ and are illuminated at $\theta_i \approx 0°$, then $E_{z2}$ will be almost singular. Likewise, we predict that the same singularity will occur even for an extremely thin EEZ slab, i.e., $h \approx 0$ nm, illuminated at $\theta_i \neq 0$. A viable way to achieve a very thin layer of EEZ metamaterial may involve the use of transformation optics techniques in order to tailor the effective optical properties of one-atom-thick materials, e.g., graphene, by properly patterning its surface [38].

b. *Total transmission condition*

The set of parameters that leads to TTC is $\theta_i = \theta_i^B$ and $\hat{\varepsilon}_2 \neq 0$, for which $\Gamma(\hat{\varepsilon}_2, \theta_i^B) = \lim_{\theta_i \to \theta_i^B} \Gamma(\hat{\varepsilon}_2, \theta_i) = 0$ and $T(\hat{\varepsilon}_2, \theta_i^B) = \lim_{\theta_i \to \theta_i^B} T(\hat{\varepsilon}_2, \theta_i) = e^{ik_{z2}h}$ which implies that $\left|T\left(\hat{\varepsilon}_2,\theta_i^B\right)\right|=1$ in the lossless case. The value of $E_{z2}$ is obtained via Eq. (10) with $\theta_i=\theta_i^B$ and $\hat{\varepsilon}_2 \neq 0$, leading to

$$E_{z2}(x,z) = -E_1 \frac{1}{\sqrt{\hat{\varepsilon}_2(1+\hat{\varepsilon}_2)}} e^{ik_1 \frac{\sqrt{\hat{\varepsilon}_2}}{\sqrt{1+\hat{\varepsilon}_2}}\left(x+z\sqrt{\hat{\varepsilon}_2}\right)}. \quad (13)$$

We stress that, for Brewster incidence, FIE is independent of the slab thickness $h$ due to the absence of any reflection form the two interfaces, i.e., $A_f = 1$ and $A_b = 0$ in Eq. (11), as can be explicitly noted by looking at Eq. (13) [equal to Eq. (7) for the two half space case to remark the thickness independence]. Moreover, the field in (13) depends inversely on $\hat{\varepsilon}_2$, thus emphasizing the importance of ENZ capabilities of medium 2 for field enhancements. The same arguments regarding the evaluation of $E_{z1}$ and $E_{z2}$ in the $(\hat{\varepsilon}_2,\theta_i)$ space detailed in Sec. 5.1b, and the $1/\sqrt{\hat{\varepsilon}_2}$ singular behavior, apply here.

  c. *Critical angle condition*

The critical angle condition (CAC) is defined by $\hat{\varepsilon}_2 \neq 0$ and $\theta_i = \theta_i^C$, which implies that $k_{z2} = 0$. This leads to

$$\Gamma\left(\hat{\varepsilon}_2,\theta_i^C\right) = \frac{ihk_{z1}\hat{\varepsilon}_2}{-ihk_{z1}\hat{\varepsilon}_2 + 2},$$
$$T\left(\hat{\varepsilon}_2,\theta_i^C\right) = \frac{2}{-ihk_{z1}\hat{\varepsilon}_2 + 2}. \quad (14)$$

It follows that $\left|T\left(\hat{\varepsilon}_2,\theta_i^C\right)\right| \approx 1$ and $\left|\Gamma\left(\hat{\varepsilon}_2,\theta_i^C\right)\right| \approx 0$ when $\left|\hat{\varepsilon}_2 k_{z1} h\right| \ll 2$, which is likely to be satisfied when considering ENZ slabs of subwavelength thickness. This result suggests that light transmission at and above the critical angle for slabs of ENZ media with finite thickness is

mediated by tunneling [4-5] of evanescent waves excited at the input interface via frustrated total internal reflection [39]. The low permittivity value of the slab tends to merge the CAC and the TTC points, as mentioned in Sec. 2. The value of the $z$ component of the field may be obtained by using Eq. (10) with $\theta_i = \theta_i^C$ and $\hat{\varepsilon}_2 \neq 0$, leading to

$$E_{z2}(x,z) = E_1 \frac{2ihk_1\sqrt{1-\hat{\varepsilon}_2}\hat{\varepsilon}_2 - 2}{\left(-ihk_1\sqrt{1-\hat{\varepsilon}_2}\hat{\varepsilon}_2 + 2\right)\sqrt{\hat{\varepsilon}_2}} e^{ik_1\sqrt{\hat{\varepsilon}_2}x}. \quad (15)$$

Again, we note an inverse dependence on $\hat{\varepsilon}_2$ for the field in (15), thus emphasizing the importance of ENZ capabilities of medium 2 for field enhancements. We note that Eq. (15) tends to Eq. (8) when $h \to +\infty$ and that arguments similar to those used in Sec. 5.1c regarding the evaluation of $E_{z1}$ and $E_{z2}$, and the $1/\sqrt{\hat{\varepsilon}_2}$ singular behavior, apply here.

*d. Representative example 1: Analysis by varying the permittivity of the ENZ slab*

As an example, we now suppose to have a TM-polarized plane wave incident at $\theta_i = 15°$ at 400 THz on a slab with thickness $h$ = 400 nm. Transmission, reflection and FIE in Eq. (3) computed at $z = 0^+$ are shown in Fig. 12 as a function of the dielectric contrast in the interval $-0.1 < \hat{\varepsilon}_2 < 0.1$. When $\hat{\varepsilon}_2 = 0$, $\Gamma(0,\theta_i) = 1$ and $T(0,\theta_i) = 0$, as predicted in Sec. 5.2a. Moreover, the FIE in Eq. (3) is about 7.6 as dictated by the limit in Eq. (12). When $\hat{\varepsilon}_2 = 0.0718$, the TTC at the Brewster angle occurs at $\theta_i^B = 15°$ [Eq. (2)], so that $\Gamma(\hat{\varepsilon}_2,\theta_i^B) = 0$ and $\left|T(\hat{\varepsilon}_2,\theta_i^B)\right| = 1$, as described in Sec. 5.2b, and the FIE $\approx 13$ as in Eq. (13). The CAC condition occurs at $\theta_i^C = 15°$ when $\hat{\varepsilon}_2 = 0.067$ [Eq. (1)] and does not inhibit light transmission, as discussed in Sec. 5.2c, leading to the FIE $\approx 15.4$ [Eq. (15)].

However, the maximum FIE does not occur either at the critical angle $\theta_i^C$, or at the Brewster angle $\theta_i^B$, but occurs at $\theta_i > \theta_i^C$. Indeed, maximum FIE is obtained when $\hat{\varepsilon}_2 = 0.0477$, for which the critical angle according to Eq. (1) is $\theta_i^C = 12.62° < \theta_i = 15°$. This effect is due to the finite thickness of the slab, which implies the presence of two interfaces free space/ENZ and ENZ/free space. As shown in Fig. 12(b), maximum FIE approaches the critical angle for increasing slab thickness $h$, consistent with the semi-infinite space results described in Sec. 5.1 (Fig. 10). This result suggests that thick ENZ slabs provide large FIE in proximity of the critical angle, and proves that the CAC condition is important even for layers of finite thickness. The FIE peaks at $\hat{\varepsilon}_2 \approx 0.09$ for the thicker slab with $h$ = 2500nm [blue dashed-dotted curve in Fig. 12(b)] is due to the first-order Fabry-Pérot resonance of the slab. Therefore, one may boost nonlinear interactions by exciting this kind of resonances in optically thick slabs, as reported in Ref. [20] for second harmonic generation in hyperbolic, low permittivity slabs. However, for isotropic ENZ slabs as those considered in the present paper, the FIE levels achieved close to the CAC condition are much higher than that at the Fabry-Pérot resonance, as displayed in Fig. 12(b) as well as Figs. 7(d) and 7(f).

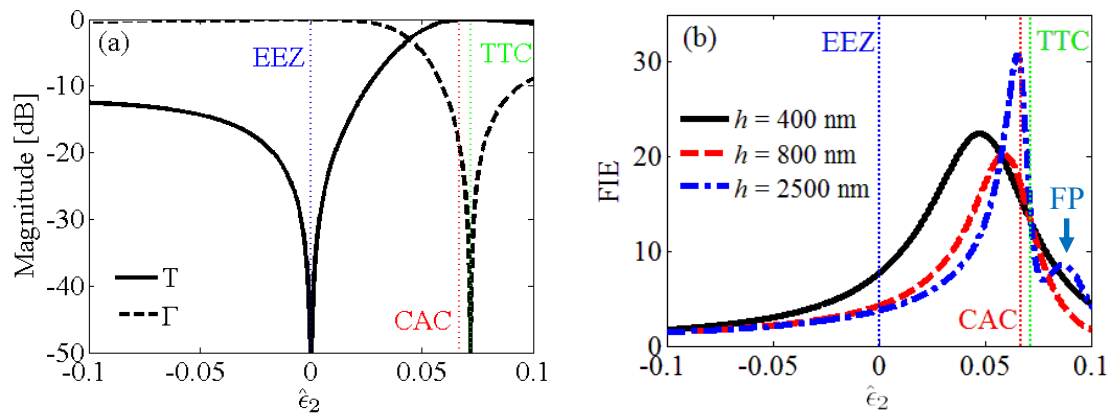

Fig. 12. (a) Magnitude of transmission and reflection coefficients and (b) the FIE in Eq. (3) computed at $z = 0^+$ as a function of the dielectric contrast $\hat{\varepsilon}_2$ assuming a TM-polarized plane wave incident at $\theta_i = 15°$ at 400 THz. Slab thickness is $h = 400$ nm for part (a). In part (b), FP indicates the location of the Fabry-Pérot resonance when $h = 2500$ nm.

*e. Representative example 2: Analysis by varying the angle of incidence*

A TM-polarized plane wave is incident with an angle in the range $0° < \theta_i < 20°$ at 400 THz. As before, we assume that slab thickness is $h = 400$ nm. Transmission and reflection coefficients are shown in Fig. 13 as a function of the incident angle $\theta_i$ for different values of the dielectric contrast $\hat{\varepsilon}_2$. When $\hat{\varepsilon}_2 = 0.1$, Eqs. (1) and (2) yield $\theta_i^B \approx 17.55°$ and $\theta_i^C = 18.44°$. Accordingly, the TTC at the Brewster angle $\theta_i^B$ is verified, and $\Gamma(\hat{\varepsilon}_2, \theta_i^B) = 0$ and $|T(\hat{\varepsilon}_2, \theta_i^B)| = 1$, as described in Sec. 5.2b. Also the CAC at $\theta_i^C$ is verified, and $|\Gamma(\hat{\varepsilon}_2, \theta_i^C)| \approx 0$ and $|T(\hat{\varepsilon}_2, \theta_i^C)| \approx 1$, as expected from Eq. (14). Similar behavior occurs for $\hat{\varepsilon}_2 = 0.01$, although now Eqs. (1) and (2) yield $\theta_i^B \approx 5.71°$ and $\theta_i^C = 5.74°$, very close in value [almost superimposed in Fig. 13(b)]. When $\hat{\varepsilon}_2 = -0.1$, Eqs. (1) and (2) yield complex angles $\theta_i^B \approx i18.76°$ and $\theta_i^C = i17.83°$. As a result TTC and CAC cannot be observed for any angle of incidence [Fig. 13(c)]. Indeed, analyzing the same situation for complex angles of incidence, i.e., using inhomogeneous waves, TTC and CAC take place [Fig. 13(d)].

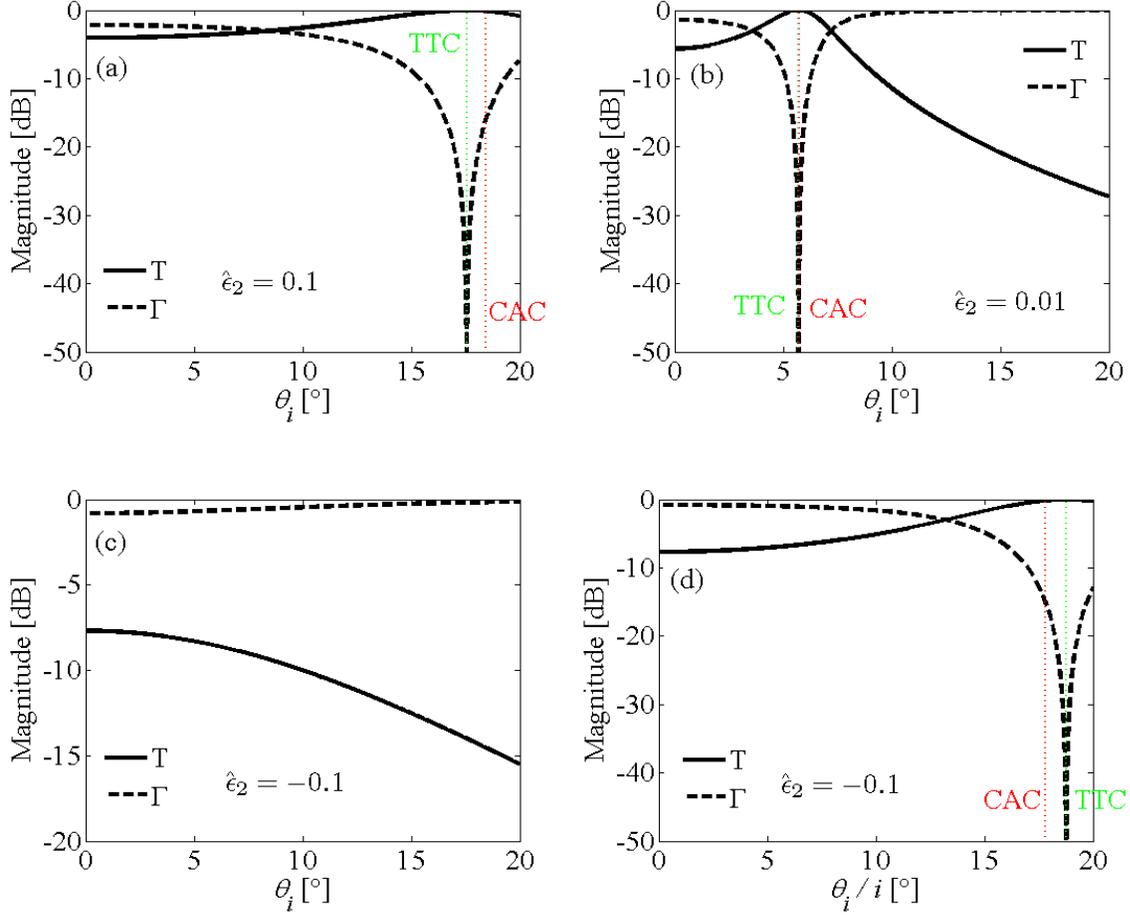

Fig. 13. Magnitude of transmission and reflection coefficients as a function of the incident angle $\theta_i$ for different values of the dielectric contrast $\hat{\varepsilon}_2$, assuming a TM-polarized plane wave incidence at 400 THz. Slab thickness is $h = 400$ nm.

In Fig. 14 we show the FIE in Eq. (3) computed at $z = 0^+$ as a function of the incident angle $\theta_i$ for the same cases analyzed in Fig. 13. The figure shows that the incident angle is pivotal to achieve large FIE. Indeed, the curve that corresponds to $\hat{\varepsilon}_2 = 0.01$ yields FIE $\approx 9.84$ at $\theta_i = 15°$, in agreement with the result shown in Fig. 12(b); the result in Fig. 14 also shows that illuminating at $\theta_i \approx 6.3° > \theta_i^C = 5.74°$ instead of 15° as in Fig. 12 leads to FIE $\approx 110$. This result motivates our analyses in Sec. 3 and Sec. 4, where we showed color maps as a function of both frequency and incident angle. Moreover, when comparing the curves for various values of $\hat{\varepsilon}_2$, the FIE angular band gets smaller as $\hat{\varepsilon}_2$ decreases.

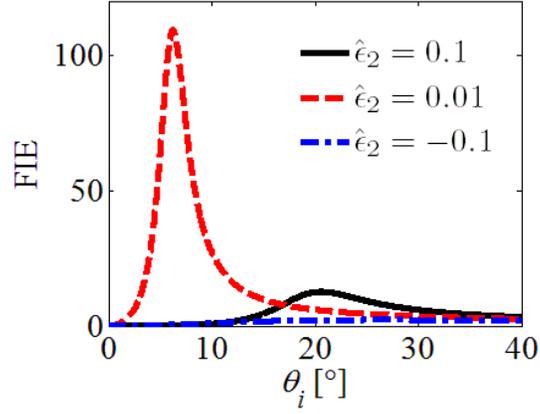

Fig. 14. The FIE in Eq. (3) computed at $z = 0^+$ as a function of the incident angle $\theta_i$ for the same cases analyzed in Fig. 13.

*f. Analysis of the effect of losses in ENZ slabs*

We now estimate the effect of losses in the ENZ slab. We assume a complex dielectric contrast $\hat{\varepsilon}_2 = \hat{\varepsilon}_{2,r} + i\hat{\varepsilon}_{2,i}$ for the 400nm-thick slab and we set $\hat{\varepsilon}_{2,r} = 0$. The absorption coefficient $A$ and the FIE in Eq. (3) calculated at $z = 0^+$ are plotted in Fig. 15 as a function of the incident angle for four different values of $\hat{\varepsilon}_{2,i}$. We observe that (i) increasing $\hat{\varepsilon}_{2,i}$ increases the incident angle of both maximum absorption and FIE; (ii) increasing $\hat{\varepsilon}_{2,i}$ broadens the absorption and FIE profiles in the angular domain; (iii) decreasing $\hat{\varepsilon}_{2,i}$ dramatically enhances the FIE inside the slab, proving our predictions of Eq. (12) in Sec. 5.2a.

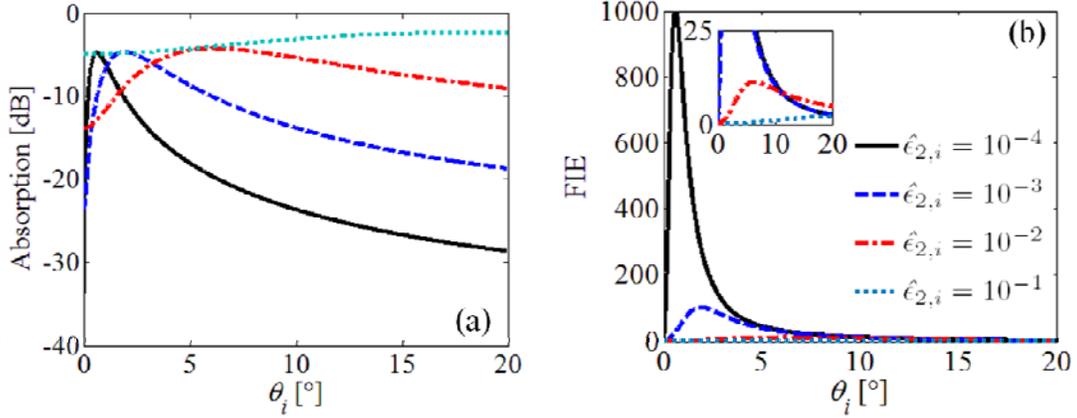

Fig. 15. (a) Magnitude of the absorption coefficient and (b) the FIE in Eq. (3) computed at $z = 0^+$ as a function of the incident angle $\theta_i$ for different values of $\hat{\varepsilon}_{2,i}$, assuming a TM-polarized plane wave incidence at 400 THz. The ENZ slab has $\hat{\varepsilon}_{2,r} = 0$ and thickness $h = 400$ nm.

## 6. Discussion: Correlation between theory and numerical results in case of gain-assisted ENZ slabs

The result in Fig. 15 justifies and supports the use of gain-assisted metamaterial ENZ slabs discussed in Sec. 4, which provided super field enhancement capabilities. We now have all the tools in place to explain the exotic behavior observed in Fig. 8. We first consider the *asymmetric* behavior of the transmission coefficient for slab 1 shown in Fig. 8(a). The fact that the value of the slab permittivity varies with frequency (Fig. 2) under oblique incidence at a fixed angle $\theta_i$ resembles the simpler theoretical analysis shown in Fig. 12. In particular, the dielectric contrast is $\hat{\varepsilon}_2 \approx (1.34 + i3.35)\times 10^{-3}$ [$\varepsilon_2 \approx (3.01 + i7.54)\times 10^{-3}$] at $f \approx$ 523 THz. Under these circumstances, Eqs. (1) and (2) yield $\theta_i^B \approx \theta_i^C = 2.85° + i1.9°$. This value is quite different from the incident angle $\theta_i = 6°$ (i.e., about the angle for which the FIE is maximized, refer to Fig. 7), thus the metamaterial is experiencing an EEZ-like condition because $\varepsilon_2 \approx 0$ and $\theta_i \neq 0°$, for which $|T(0,\theta_i)| = 0$ and $|\Gamma(0,\theta_i)| = 1$, as shown in Sec. 5.2a. At $f \approx 523.7$ THz, instead,

$\hat{\varepsilon}_2 \approx (11.1 + i2.8) \times 10^{-3}$ [$\varepsilon_2 \approx (25 + i6.3) \times 10^{-3}$]. For this value Eqs. (1) and (2) yield $\theta_i^B \approx \theta_i^C = 6.1° + i0.75°$, similar to $\theta_i = 6°$. Thus, losses are low enough to allow TTC/CAC-like conditions to occur, the transmission $\left|T(\hat{\varepsilon}_2, \theta_i^{B,C})\right|$ to peak and the reflection $\left|\Gamma(\hat{\varepsilon}_2, \theta_i^{B,C})\right|$ to drop even for an incident homogeneous plane wave.

A similar explanation can be provided for slab 2 in Fig. 8(b). The slab dielectric contrast is $\hat{\varepsilon}_2 \approx (1.63 + i0.47) \times 10^{-3}$ [$\varepsilon_2 \approx (3.659 + i1.056) \times 10^{-3}$] at $f \approx 421.8$ THz. For these conditions Eqs. (1) and (2) yield $\theta_i^B \approx \theta_i^C = 2.33° + i0.33°$. This value is different from the incident angle $\theta_i = 3°$ (i.e., about the angle for which the FIE is maximized, refer to Fig. 7), thus the metamaterial is experiencing an EEZ-like condition because $\varepsilon_2 \approx 0$ and $\theta_i \neq 0°$, for which $|T(0, \theta_i)| = 0$ and $|\Gamma(0, \theta_i)| = 1$. At $f \approx 421.85$ THz, instead, $\hat{\varepsilon}_2 \approx (2.89 + i0.4) \times 10^{-3}$ [$\varepsilon_2 \approx (6.5 + i0.9) \times 10^{-3}$], which leads to $\theta_i^B \approx \theta_i^C = 3.09° + i0.2°$, close to $\theta_i = 3°$. Losses are thus low enough to allow TTC/CAC-like conditions to take place. In this case however, a smaller effective permittivity imaginary part now leads to $\left|T(\hat{\varepsilon}_2, \theta_i^{B,C})\right|$ closer to 1 and $\left|\Gamma(\hat{\varepsilon}_2, \theta_i^{B,C})\right|$ closer to 0 than the correspondent curves of the case in Fig. 8(a).

In summary, the presence of EEZ, TTC and CAC conditions leads to the peculiar behavior of the linear properties of the metamaterial slabs discussed in Sec. 4, inducing an absorption peak and, more importantly, a large boost of the electric field inside the slab. Although the gain-assisted tunneling we have discussed is extremely selective in both angular and frequency domains and its spectral shape resembles a Fano-like resonance typical of systems with electromagnetic-induced-transparency, we point out that this tunneling effect has a non-resonant nature. The selectivity is due to the simultaneous presence of a Brewster-like condition at the interface, the zero-crossing point of the slab permittivity's real part, and the damping compensation provided by the gain medium.

## 7. Conclusion and final remarks

We have investigated transmission, reflection and absorption coefficients, as well as local field enhancement, in subwavelength ENZ slabs illuminated by TM-polarized plane waves. We have analyzed various configurations, including metamaterial implementations, without and with the introduction of gain in the system. While the former is strictly dependent on material properties, the latter leads to super field enhancement in a very narrow frequency band and for specific incident angles. Our study thus shows how control of the slab permittivity enables the field to take on very large values and thus improve applications where large fields are required. We have demonstrated that the FIE in the case of a single interface may assume large values at the critical angle and is singular only under TTC and CAC conditions. A finite thickness, subwavelength ENZ slab may also exhibit very large FIE values: besides a singularity similar to the single interface case at TTC and CAC conditions, FIE is singular also in the limit for vanishing permittivity $\hat{\varepsilon}_2$ and vanishing incident angle that however has to be larger than the critical angle (also vanishing), as predicted by Eq. (12). Interestingly, the finite subwavelength thickness of the ENZ slab helps in establishing a FIE enhancement. Moreover, we predict that if damping is virtually compensated [i.e., if $\text{Im}(\hat{\varepsilon}_2) \approx 0$] near the zero-crossing point of the real part of the effective permittivity, any increases in field enhancement may be exploited to significantly lower the threshold of nonlinear processes, such as optical switching and bistability, and dramatically increase the frequency conversion efficiency in devices for the generation of coherent light sources in the UV and extreme-UV.

**Acknowledgement**

The authors associated with the University of California Irvine thank Ansys Inc. for providing HFSS that was instrumental in this analysis. This research was performed while the authors M. A. Vincenti and D. de Ceglia held a National Research Council Research Associateship award at the U. S. Army Aviation and Missile Research Development and Engineering Center.

**References**

[1] N. Garcia, E. V. Ponizovskaya, and J. Q. Xiao, "Zero permittivity materials: Band gaps at the visible," *Applied Physics Letters,* vol. 80, pp. 1120-1122, 2002.

[2] A. Alu, N. Engheta, A. Erentok, and R. W. Ziolkowski, "Single-Negative, Double-Negative, and Low-index Metamaterials and their Electromagnetic Applications," *Antennas and Propagation Magazine, IEEE,* vol. 49, pp. 23-36, 2007.

[3] J. Brown, "Artificial dielectrics having refractive indices less than unity," *Proceedings of the IEE - Part III: Radio and Communication Engineering,* vol. 100, pp. 319-320, 1953.

[4] M. Silveirinha and N. Engheta, "Tunneling of Electromagnetic Energy through Subwavelength Channels and Bends using ε-Near-Zero Materials," *Physical Review Letters,* vol. 97, p. 157403, 2006.

[5] R. Liu, Q. Cheng, T. Hand, J. J. Mock, T. J. Cui, S. A. Cummer, and D. R. Smith, "Experimental Demonstration of Electromagnetic Tunneling Through an Epsilon-Near-Zero Metamaterial at Microwave Frequencies," *Physical Review Letters,* vol. 100, p. 023903, 2008.

[6] K. C. Gupta, "Narrow-beam antennas using an artificial dielectric medium with permittivity less than unity," *Electronics Letters,* vol. 7, pp. 16-18, 1971.

[7] I. Bahl and K. Gupta, "A leaky-wave antenna using an artificial dielectric medium," *Antennas and Propagation, IEEE Transactions on,* vol. 22, pp. 119-122, 1974.

[8] S. Enoch, G. Tayeb, P. Sabouroux, N. Guerin, and P. Vincent, "A metamaterial for directive emission," *Physical Review Letters,* vol. 89, p. 213902, Nov 2002.

[9] G. Lovat, P. Burghignoli, F. Capolino, D. R. Jackson, and D. R. Wilton, "Analysis of directive radiation from a line source in a metamaterial slab with low permittivity," *Ieee Transactions on Antennas and Propagation,* vol. 54, pp. 1017-1030, Mar 2006.

[10] P. Burghignoli, G. Lovat, F. Capolino, D. R. Jackson, and D. R. Wilton, "Directive Leaky-Wave Radiation From a Dipole Source in a Wire-Medium Slab," *Antennas and Propagation, IEEE Transactions on,* vol. 56, pp. 1329-1339, 2008.

[11] G. Lovat, P. Burghignoli, F. Capolino, and D. R. Jackson, "High directivity in low-permittivity metamaterial slabs: Ray-optic vs. leaky-wave models," *Microwave and Optical Technology Letters,* vol. 48, pp. 2542-2548, 2006.

[12] A. Alù, M. G. Silveirinha, A. Salandrino, and N. Engheta, "Epsilon-near-zero metamaterials and electromagnetic sources: Tailoring the radiation phase pattern," *Physical Review B,* vol. 75, p. 155410, 2007.

[13] N. Engheta, "Circuits with Light at Nanoscales: Optical Nanocircuits Inspired by Metamaterials," *Science,* vol. 317, pp. 1698-1702, September 21, 2007 2007.

[14] M. Navarro-Cía, M. Beruete, I. Campillo, and M. Sorolla, "Enhanced lens by ε and μ near-zero metamaterial boosted by extraordinary optical transmission," *Physical Review B,* vol. 83, p. 115112, 2011.

[15] B. Edwards, A. Alù, M. G. Silveirinha, and N. Engheta, "Experimental Verification of Plasmonic Cloaking at Microwave Frequencies with Metamaterials," *Physical Review Letters,* vol. 103, p. 153901, 2009.

[16] A. Monti, F. Bilotti, A. Toscano, and L. Vegni, "Possible implementation of epsilon-near-zero metamaterials working at optical frequencies," *Optics Communications,* vol. 285, pp. 3412-3418, 2012.

[17] C. Argyropoulos, P.-Y. Chen, G. D'Aguanno, N. Engheta, and A. Alù, "Boosting optical nonlinearities in ε-near-zero plasmonic channels," *Physical Review B,* vol. 85, p. 045129, 2012.

[18] A. Ciattoni, C. Rizza, and E. Palange, "Extreme nonlinear electrodynamics in metamaterials with very small linear dielectric permittivity," *Physical Review A,* vol. 81, p. 043839, 2010.

[19] M. A. Vincenti, D. de Ceglia, A. Ciattoni, and M. Scalora, "Singularity-driven second- and third-harmonic generation at ε-near-zero crossing points," *Physical Review A,* vol. 84, p. 063826, 2011.

[20] A. Ciattoni and E. Spinozzi, "Efficient second-harmonic generation in micrometer-thick slabs with indefinite permittivity," *Physical Review A,* vol. 85, p. 043806, 2012.

[21] J. A. Gordon and R. W. Ziolkowski, "The design and simulated performance of a coated nano-particle laser," *Opt. Express,* vol. 15, pp. 2622-2653, 2007.

[22] J. A. Gordon and R. W. Ziolkowski, "CNP optical metamaterials," *Opt. Express,* vol. 16, pp. 6692-6716, 2008.

[23] S. Campione, M. Albani, and F. Capolino, "Complex modes and near-zero permittivity in 3D arrays of plasmonic nanoshells: loss compensation using gain [Invited]," *Opt. Mater. Express,* vol. 1, pp. 1077-1089, 2011.

[24] S. Campione and F. Capolino, "Composite material made of plasmonic nanoshells with quantum dot cores: loss-compensation and ε-near-zero physical properties," *Nanotechnology,* vol. 23, p. 235703, 2012.

[25] A. Ciattoni, R. Marinelli, C. Rizza, and E. Palange, "|\epsilon|-Near-zero materials in the near-infrared," *Applied Physics B,* pp. 1-4, 2012/10/01 2012.

[26] P. B. Johnson and R. W. Christy, "Optical Constants of the Noble Metals," *Physical Review B,* vol. 6, p. 4370, 1972.

[27] P. R. West, S. Ishii, G. V. Naik, N. K. Emani, V. M. Shalaev, and A. Boltasseva, "Searching for better plasmonic materials," *Laser & Photonics Reviews,* vol. 4, pp. 795-808, Nov 2010.

[28] G. V. Naik, J. Kim, and A. Boltasseva, "Oxides and nitrides as alternative plasmonic materials in the optical range [Invited]," *Opt. Mater. Express,* vol. 1, pp. 1090-1099, 2011.

[29] A. Boltasseva and H. A. Atwater, "Low-Loss Plasmonic Metamaterials," *Science,* vol. 331, pp. 290-291, January 21, 2011 2011.

[30] G. Ohman, "The pseudo-Brewster angle," *Antennas and Propagation, IEEE Transactions on,* vol. 25, pp. 903-904, 1977.

[31] S. P. F. Humphreys-Owen, "Comparison of Reflection Methods for Measuring Optical Constants without Polarimetric Analysis, and Proposal for New Methods based on the Brewster Angle," *Proceedings of the Physical Society,* vol. 77, p. 949, 1961.

[32] S. Campione, S. Steshenko, M. Albani, and F. Capolino, "Complex modes and effective refractive index in 3D periodic arrays of plasmonic nanospheres," *Opt. Express,* vol. 19, pp. 26027-26043, 2011.

[33] E. Dulkeith, A. C. Morteani, T. Niedereichholz, T. A. Klar, J. Feldmann, S. A. Levi, F. C. J. M. van Veggel, D. N. Reinhoudt, M. Möller, and D. I. Gittins, "Fluorescence Quenching of Dye Molecules near Gold Nanoparticles: Radiative and Nonradiative Effects," *Physical Review Letters,* vol. 89, p. 203002, 2002.

[34] A. Ciattoni, R. Marinelli, C. Rizza, and E. Palange, "|ε|-Near-Zero materials in the near-infrared," *arxiv:1107.5540,* 2011.

[35] M. A. Vincenti, S. Campione, D. d. Ceglia, F. Capolino, and M. Scalora, "Gain-assisted harmonic generation in near-zero permittivity metamaterials made of plasmonic nanoshells," *New Journal of Physics,* vol. 14, p. 103016, 2012.

[36] D. M. Pozar, *Microwave Engineering*, 3rd ed. New York: Wiley, 2005.

[37] M. Born and E. Wolf, *Principles of Optics*, 7th ed. Cambridge: University Press, 2002.

[38] A. Vakil and N. Engheta, "One-Atom-Thick Metamaterials and Transformation Optics with Graphene," *Opt. Photon. News,* vol. 22, pp. 44-44, 2011.

[39] S. Zhu, A. W. Yu, D. Hawley, and R. Roy, "Frustrated total internal reflection: A demonstration and review," *American Journal of Physics,* vol. 54, pp. 601-607, 1986.